\documentstyle[a4,11pt,epic,eepic]{article}
\date{}

\newcommand{\eqref}[1]{(\ref{#1})}

\title{Existence of Constant Mean Curvature Hypersurfaces 
in Asymptotically Flat Spacetimes}

\author{Lars Andersson\footnotemark{$^*$} \\
Department of Mathematics \\
Royal Institute of Technology \\
S-100 44 Stockholm, Sweden \\
{\tt larsa\char'100math.kth.se} \\
\and
Mirta S. Iriondo\footnotemark{$^\dagger$} \\ 
FaMAF \\
Medina Allende y Haya de la Torre \\
Ciudad Universitaria \\
5000 C\'ordoba, Argentina \\
{\tt mirta\char'100fis.uncor.edu}
}

\newtheorem{theorem}{Theorem}[section]
\newtheorem{lemma}[theorem]{Lemma}

\newtheorem{remark}{Remark}[section]
\newtheorem{definition}{Definition}[section]

\newcommand{\Scri}{{\cal I}}

\newcommand{\new}{\mbox{\tt\char'134begin\{New\}}}   

\def\endnew{\mbox{\tt\char'134end\{New\}}}  
\newcommand{\lars}{{\mbox{\bf *LA*}}}
\newcommand{\mirta}{{\mbox{\bf *MI*}}}
\newcommand{\mnote}[1]{\marginpar{\raggedright\footnotesize\em #1}}
\newcommand{\DELETENOTES}{ 
\renewcommand{\mnote}[1]{}
\renewcommand{\lars}{}
\renewcommand{\mirta}{}
\renewcommand{\new}{}
\renewcommand{\endnew}{}
}

\newcommand{\text}[1]{\mbox{#1}}
\newcommand{\qed}{$\Box$}
\newenvironment{pf}{{\bf Proof:}}{\hfill{\bf\qed}\smallskip\\}
\newcommand{\tfrac}[2]{\frac{#1}{#2}}
\newcommand{\dfrac}[2]{\frac{#1}{#2}}

\renewcommand{\Re}{{\Bbb R}}
\renewcommand{\Re}{{\bf R}}
\DELETENOTES  

\begin{document}
\maketitle

\def\thefootnote{}
\footnotetext{
{\em \textup{1991} Mathematics Subject Classification.} 
Primary: 53A10 
Secondary: 53B30 
}

\footnotetext{{\em  Key words and phrases.} 
Constant mean curvature, Lorentzian geometry, asymptotic structure}

\footnotetext{\footnotemark{$^*$} Visiting Scientist: Max--Planck--Institut f\"ur Gravitationsphysik (Albert--Einstein--Institut) Schlaatzweg 1, D-14473 Potsdam, Germany. Supported in part by NFR, contract no. F-FU
4873-307.}
\footnotetext{\footnotemark{$^\dagger$}Supported by STINT, The Swedish Foundation for International Cooperation in Research and Higher Education}

\begin{abstract}
\noindent
The problem of existence of spacelike hypersurfaces with constant mean 
curvature in asymptotically flat spacetimes is considered for a  class of 
asymptotically Schwarzschild spacetimes satisfying an interior condition. 
Using a barrier construction, a proof is given of the existence of 
complete hypersurfaces with constant mean cuvature  which intersect null 
infinity in a regular cut. 
\end{abstract}


\thispagestyle{empty}


\newpage
\setcounter{page}{1}

\section{Introduction}
In this paper we prove existence of complete  constant mean curvature
(CMC) hypersurfaces asymptotic to a cut of null infinity, $\Scri^+$,
in a class of asymptotically flat spacetimes.

Entire CMC hypersurfaces in Minkowski space have been classified
by Treibergs in \cite{Treiberg} where the existence of a CMC
hypersurface asymptotic to any $C^2$ cut of  $\Scri^+$
was proved.  
The method used in \cite{Treiberg} is based on the ideas of 
\cite{Che-Yau} and is limited to the case when the ambient space is
flat.

R. Bartnik and L. Simon \cite{BS} have studied   the Dirichlet
problem for the prescribed mean curvature equation in the case of
Minkowski space.
The existence and regularity of hypersurfaces with
prescribed mean curvature follow once a uniform gradient bound is
established. The first apriori gradient 
estimate in non-flat spacetime was given by  C. Gerhardt \cite{Ger}.

R. Bartnik  has extended the existence and regularity results for the
Dirichlet problem to Lorentzian manifolds   in \cite{Bartnik2} and   
\cite{Bartnik3}, and in
\cite{Bartnik} he proved the existence of maximal hypersurfaces asymptotic to spatial infinity in
asymptotically flat spacetimes  at spatial infinity, satisfying an
{\em interior condition} which is a uniform bound on the size of domains of
dependence. 

In this paper we prove existence of complete CMC hypersurfaces asymptotic to a 
cut of $\Scri^+$ in a class of {\em asymptotically Schwarzschild
spacetimes} (Definition \ref{def:asymp-schwarz}).
%
\mnote{\lars small change here}
The method used in this paper is similar to that used in \cite{Bartnik}, 
i.e. we prove existence of a CMC hypersurface by proving convergence of 
the solutions to a sequence of Dirichlet problems, using barriers at infinity
and a  condition on the causal structure. 
In the case studied in this paper, the {\em interior condition } used
in \cite{Bartnik} may be replaced by a {\em future interior condition},
see Definition \ref{def:hypprop}. 
\mnote{put in remark about regularity??}

\subsection{Overview of this paper}
\label{sec:overview}

In this paper, in order to keep calculations and notation manageable 
we deal only with 4-dimensional spacetimes and with mean curvature $k=3$.
The results generalize in a straightforward way to general dimension and
arbitrary constant mean curvature.

In Section \ref{sec:prelim} we define asymptotically Schwarzschild
spacetimes, (see Definition \ref{def:asymp-schwarz}). These spacetimes admit
a future null infinity $\Scri^+$ as an asymptote with conformal factor
$\Omega$ and coordinates $(u,x,y^A)$ in a neighborhood ${\cal N}$ of
$\Scri^+$ so that on $\Scri^+ = \{x=0\}$, $u$ is the affine parameter of
null geodesics, $\tilde \nabla u \big{|}_{\Scri^+} = \partial_x$,
$\partial_x$ is null on ${\cal N}$ and $y^A$ are coordinates on $S^2$. 
It may be shown that any asymptotically flat and empty spacetime admits such
a structure locally near $\Scri^+$ if the stress energy tensor decays to 
second order at $\Scri^+$ \cite[Lemma 4.3]{iriondo}. 

In Section \ref{sec:barriers} we construct in terms of $(u,x,y^A)$ a
foliation of spacelike hypersurfaces $\{S_{\tau}\}_{\tau \in (0,\infty)}$
and prove that this foliation has properties analogous to those of the
hyperboloids in Minkowski space. In particlular $\{S_{\tau}\}$ can be used
as upper and lower barriers for CMC hypersurfaces near $\Scri^+$ and
further, the apriori estimates given by \cite[Theorem 3.1]{Bartnik}
apply.

In Section \ref{sec:existence}, Theorem \ref{thm:main}, we prove our main 
result,
namely the existence of complete $k=3$ CMC hypersurfaces in asymptotically
Schwarzschild spacetimes satisfying the {\em future interior condition} 
(Definition \ref{def:hypprop}). This condition is a uniform
bound on the size of domains of dependence in the future of a reference
hypersurface together with the condition that domains of dependence of
compact sets be compact. 

Using the maximum principle, the future interior condition and the
properties of $\{S_{\tau}\}$ proved in Section \ref{sec:barriers}, we
use an integral estimate to
construct a global upper barrier for $k=3$ hypersurfaces, see Theorem
\ref{thm:fund-est}. 

Using causality and the global upper barrier we prove using an Arzela-Ascoli
type argument that the solutions to a sequence of Dirichlet problems has a
subsequence which converges to a spacelike hypersurface $M$ with mean
curvature $k_M=3$.
Using $\{S_\tau \}$ as lower barriers for small $\tau$ we prove using a
gradient estimate that the mean curvature operator is uniformly elliptic on
$M$ which allows us to prove that $M$ is complete. 

\noindent{\bf Acknowledgements:} We are grateful to Robert Bartnik for
helpful remarks.

\vspace*{.2in}

\section{Preliminaries}
\label{sec:prelim}
The aim of this section is to present definitions and some simple calculations
for later reference.  

\subsection{Notation}
\label{sec:not-basic}
A spacetime $( V, g)$ is a smooth four-dimensional manifold with a smooth Lorentzian  metric with
signature $(-,+,+,+)$. We denote the metric pairing by $\langle\cdot,\cdot\rangle$, the canonical connection
by  $\nabla$, and use the summation convention with index ranges 
$0\leq \alpha , \beta \leq 3,\;1\leq i, j\leq 3$ where we use lower case 
latin letters $i,j,k \dots$ for
spacelike frame indices. Finally we use  upper case latin letters $A,B,C\dots$ for
indices with the range $2,3$.
We shall use the notation of Hawking and Ellis \cite{HE} in describing 
causal relationships.

We suppose that $V$ is time-orientable and that $T_1$ is a $C^2$ unit timelike 
vector field on $V$. Let  $T^*_1$ be  the dual of $T_1$ with respect the 
metric (in local coordinates $T^*_1{}_\mu = g_{\mu\nu}T_1^{\nu}$) and  
construct a reference Riemannian metric 
\begin{equation}\label{eq:gEdef}
g_E= g+2 T^*_1\otimes T^*_1. 
\end{equation}
This metric will be  used to measure the size of tensors and their covariant derivatives. For any tensor $B$, we define the norms
\begin{eqnarray*}
\Vert B\Vert (x) &=& (g_E(B,B)(x))^{1/2},\quad x\in V, \\
\Vert B\Vert &=& \sup\lbrace \Vert B\Vert (x): x \in V\rbrace, \\
\Vert B\Vert_k &=&\sum\limits_{j=0}^k \Vert\nabla^j  B\Vert.
\end{eqnarray*}
In the estimates we will use $C$ to denote a generic positive constant.

Assume that $(V,g)$ is stably causal, let $t$ be a $C^{\infty}$ time 
function on $V$ and let $S_t$ denote the level sets of $t$. 
The foliation $\{S_t\}$ will be called the reference foliation and the 
$S_t$ will be called reference slices.

Using local coordinates $(x^\mu) = (t,x^i)$ where $(x^i)$ are coordinates on the slices
$S_t$, the metric $g$ takes the form
$$
g_{\mu\nu} dx^\mu dx^\nu = 
-(\alpha^2 - \beta^2)dt^2 + 2\beta_i dx^i dt + \gamma_{ij} dx^i dx^j,
$$
where $\gamma$ is the induced metric on $S_t$, $\alpha$ is the lapse
function and $\beta$ is the shift vector.

We write $\partial _i,\,\partial_t$ for coordinate tangent vectors and denote 
partial derivatives by subscripts,
so the tangential gradient operator on the slices $S_t$ is 
$D\phi = \gamma^{ij} \phi_j\partial_i$ and the metric pairing 
$\gamma (X,Y)= X\cdot Y,$ where  $X, Y\in TS_t$. Then the future-directed 
unit normal $\hat T$ of $S_t$ can be written as
\begin{eqnarray}
\hat T&=& -\alpha  \nabla  t \nonumber \\
&=&  \alpha^{-1}(\partial_t -\beta) .
\label{eq:T} 
\end{eqnarray}
We will work in the setup used by Bartnik in \cite{Bartnik2} and
\cite{Bartnik}. In particular we will often work with the mean curvature of a
hypersurface $M$ expressed in terms of a reference foliation $S_t$ and a
height function $w$. Note that we use 
the convention that objects defined w.r.t. $M$ are distinguished by an upper or
lower index $M$, but objects defined w.r.t. $S_t$ are not given an upper
index $0$ as in \cite{Bartnik}. In particular, the notation $k(w)$, where $w$
is a height function, denotes the mean curvature operator defined w.r.t the
reference foliation operating on $w$.

Choosing an orthonormal frame $\{ z_i\}_{i=1}^3$ on $S_t$ yields an adapted 
orthonormal frame $\{ z_\mu\}_{\mu=0}^3$ on $V$, where $z_0=\hat T$. The 
divergence operator, the second fundamental and 
the mean curvature for the reference slices are given by
\begin{eqnarray*}
 \text{div}X &=& \sum_{i=1}^3 
\langle z_i,\nabla_{z_i} X\rangle, \quad X\in\Gamma (TV)\\
 k(z_i,z_j) &=&\langle z_i,\nabla_{z_j} \hat T\rangle ,\\
 k &=& \sum_{i=1}^3 k(z_i,z_i)=\text{div} \hat T.
\end{eqnarray*}
For a spacelike hypersurface $M \subset V$, we define
\begin{eqnarray*}
\text{div}_M X &=& \sum_{i=1}^3 \langle e_i,\nabla_{e_i} X\rangle, \quad X\in\Gamma (TV),
\\
\nabla^M \phi &=& \sum_{i=1}^3  e_i (\phi) e_i, \quad\phi \in C^\infty (M), 
\end{eqnarray*}
where $\lbrace e_i\rbrace^3_{i=1} $ is any orthonormal frame on $M$. 

Let $N$ be the future-directed unit normal to $M$. Then the second fundamental 
form  and mean curvature of $M$ are 
\begin{eqnarray*}
k_M(e_i,e_j) &=&\langle e_i,\nabla_{e_j} N\rangle ,\\
k_M &=& k(e_i,e_i)=\text{div}_M  N.
\end{eqnarray*}

The height function $w$ of $M$ is defined as the restriction of the time 
function to $M$, i.e.  $w=t|_{M}$ and extended to $V$ by the requirement
$\partial_t w= 0$. Thus $M$ is a level set of $\Phi=t-w$ and with 
$\eta^{-2} = - \langle \nabla \Phi, \nabla \Phi \rangle$,  
$N$ can be written as 
\begin{eqnarray}
N &=& -\eta\nabla \Phi \label{eq:N0} \\
 &=& \nu (W+\hat T), \label{eq:N}
\end{eqnarray}
where $W=(1+\beta\cdot Dw)^{-1} \alpha Dw$ and
$\nu = -\langle\hat T,N\rangle  = (1-|W|^2)^{-1/2}$.

In the proof of Theorem \ref{thm:main} we will need  adapted frames
on $TM$ and $TS$.
On  $TM$  we choose the frame $[e_1,\lbrace e_A\rbrace  _{A=2,3}]$, where
\begin {equation}
e_1 = |\nabla^M w|^{-1}\nabla^M w,\quad \text{if } |\nabla^M w|\neq 0,
\label{eq:framee}
\end{equation}
and $e_A$ are two common tangent vectors of $TM$ and $TS_t$ at the
intersection
point, the frame on $TS_t$ will be chosen as
$\lbrace z_i\rbrace^3 _{i=1}= [z_1,\lbrace e_A\rbrace _{A=2,3}]$,
where
\begin{equation}
z_1 = |Dw |^{-1} Dw, \quad\text{if } |Dw|\neq 0.
\label{eq:framev}
\end{equation}
$e_1$ and $z_1$ are related by 
\begin{equation}
\label{eq:framerel}
e_1 = \nu ( z_1 + |W| \hat T), 
\end{equation}
cf. \cite[p. 159]{Bartnik}.

We end this Subsection by stating a formula which will be used in Section
\ref{sec:existence}.
From the definitions, we have
\begin{eqnarray*}
\alpha\nabla^Mw&=&\alpha(\nabla t)^\Vert\\
&=&\nu N-\hat T.
\end{eqnarray*}
Applying the divergence operator to the last equation and using 
$$
\text{div}_M(\hat T) =k+\nu^2\langle W,\nabla_{W}\hat T\rangle 
+\nu^2\langle W,\nabla_{\hat T}\hat T \rangle ,
$$
cf. \cite[Equation (2.18)]{Bartnik}, we obtain
\begin{eqnarray}
k_M\nu&=&\text{div}_M(\alpha\nabla^M w)+\nonumber\\
& &+k+(\nu^2-1)\big{(}k(\hat W, \hat W) -\alpha^{-1}\hat T(\alpha )\big{)}
+\langle \nabla^M w,\nabla^M \alpha\rangle,
\label{eq:mcf}
\end{eqnarray}
where $\hat W = W/|W|$.

\subsection{Asymptotically Schwarzschild spacetimes}
\label{sec:compac}
In this Section we will define asymptotically Schwarzschild spacetimes, which
are the class of spacetimes covered by the existence Theorem for CMC
hypersurfaces proved in Section \ref{sec:existence}.

\newcommand{\LOCal}{} 
Let $(V,g)$ be a spacetime with future null infinity 
$\Scri^+ \approx \Re \times S^2$ as an 
asymptote, with conformal factor $\Omega$ and unphysical metric 
$\tilde g_{\mu\nu} = \Omega^2 g_{\mu\nu}$, cf.  \cite{Geroch}. 
If the stress energy tensor decays to second order at $\Scri^+$,
it is possible to prove, cf. \cite[Lemma 4.3]{iriondo},  that there are coordinates $\{u,x,y^A\}_{A=2,3}$ 
locally at $\Scri^+$, so that $\Omega= x+{\LOCal O}(x^3)$ and 
$\tilde g_{\mu\nu}$ takes the form 
\begin{equation}
(\tilde g_{\mu\nu})=\left(\begin{array}{cccc}
-x^2 f  \;& \;1 \;  & {\LOCal O}(x^2) &{\LOCal O}(x^2)\\
1                 & 0          & 0         & 0\\
{\LOCal O}(x^2)     &0           &           & \\
{\LOCal O}(x^2)     &0           &
\multicolumn{2}{c}{\smash{\raise6pt\hbox{$\tilde g_{AB}$}}}
         \end{array}     \right),
\label{eq:goodmetr}
\end{equation}
where  $f$ is a smooth function on $\tilde V = V \cup \Scri^+$ and for 
fixed $(x,u)$, 
$\tilde g_{AB}$ is a smooth Riemannian metric on $S^2$. 

In the remainder of the paper, we will
deal with uniform estimates near $\Scri^+$. To this end we make the following
definition. 
\begin{definition} \label{def:ordodef}
Given coordinates $x^{\mu}= (u,x,y^A)$, $A=2,3$,  in  ${\cal N}$, where
$$
{\cal N}=\Re \times  [0,x_0) \times S^2 , 
$$
we say that a function $f$ on ${\cal N}$ satisfies 
$$
f ={\cal O}(x^{\alpha})
$$
for some $\alpha \in \Re$ if for some constant $C > 0$,
$$
| f (u,x,y^A) | < Cx^{\alpha} \qquad \text{for all}\qquad (u,x,y^A).
$$
We extend this definition to tensors by working with components in a given
frame.
\end{definition}  

Let $m >0$ be given.
The Schwarzchild metric with mass $m$ can be put on the form 
$x^{-2} \tilde g^{(0)}_{\mu\nu}$ where
the conformal Schwarzschild metric $\tilde g^{(0)}_{\mu\nu}$ is given by 
\begin{equation}
\tilde g^{(0)}_{\mu\nu}dx^\nu dx^\mu= -x^2h(x)du^2+2dudx
+d\theta ^2+\sin \theta^2d\phi^2 .
\label{eq:hatgdef}
\end{equation}
Here 
\begin{equation} \label{eq:h(x)def}
h(x) = 1-2mx
\end{equation}
and $y^A= (\theta, \phi)$. The unphysical Schwarzschild metric $\tilde
g^{(0)}_{\mu\nu}$ is an example of a metric of the form 
\eqref{eq:goodmetr} on a domain
of the form given in Definition \ref{def:ordodef}.

We are now ready to introduce the precise notion of asymptotic behavior at
$\Scri^+$ which we will work with in the rest of the paper. 

\begin{definition}
\label{def:asymp-schwarz}
Let $(V,g)$ be a stably causal spacetime with an asymptote $\Scri^+$, with
$\tilde V = V \cup \Scri^+$ the unphysical spacetime, 
$\Omega \in C^{\infty}(\tilde V)$ the conformal factor and
$\tilde g_{\mu\nu} = \Omega^2 g_{\mu\nu}$ the unphysical metric. 
Let $m > 0$ be given and let $x_0 < 1/2m$. 
Then $(V,g)$ is said to be asymptotically Schwarzschild if the following holds. 
\begin{enumerate}
\item 
There is a 
coordinate system $(u,x,y^A)$ on a domain ${\cal N}$ as
in Definition \ref{def:ordodef}, and an embedding $F: {\cal N} \to \tilde V$
such that in the coordinates $(u,x,y^A)$ on $F({\cal N})$,  
$\tilde g_{\mu\nu}$ 
takes the form 
\eqref{eq:goodmetr} 
with $\tilde g_{0A} = {\cal O}(x^2)$ in the sense of Definition 
\ref{def:ordodef}
and $\Lambda ^{-1} \tilde g^0_{AB} \xi^A \xi^B \leq \tilde g_{AB} \xi^A
\xi^B \leq \Lambda \tilde g^{(0)}_{AB} \xi^A \xi^B$ for some constant $\Lambda >0$
and all $\xi^A$, where $\tilde g^{(0)}_{AB}$ is the standard metric on
$S^2$.
\item $F({\cal N} \cap \{x=0\}) = \Scri^+$ and on $F({\cal N})$, $\partial_x$
is future oriented. 
\item
On $F({\cal N})$ the conditions
\begin{eqnarray}
\Omega^{-6}|\delta \tilde g_{uu}|
+\Omega^{-5}\sum_{\mu=0}^3|\partial_\mu \delta \tilde g_{uu} |
+\Omega^{-4}\sum_{\mu,\nu=0}^3
|\partial_{\mu}\partial_{\nu} \delta \tilde g_{uu} | &<&  C ,
\label{eq:asest1} \\
|\tilde g_{uA}| \Omega^{-3}+\Omega^{-2}\sum_{\mu=0}^3|\partial_\mu\tilde
g_{uA}|+
\Omega^{-1}\sum_{\mu,\nu=0}^3|\partial_{\mu}\partial_{\nu}\tilde g_{uA}| 
&<& C,  \label{eq:asest2} \\
\Omega^{-2}|\delta \tilde g_{AB} |
+\Omega^{-1}\sum_{\mu=0}^3|\partial_\mu \delta \tilde g_{AB}  |
+\sum_{\mu,\nu=0}^3|\partial_\mu\partial_{\nu} \delta \tilde g_{AB}  | &<& C,
\label{eq:asest3} 
\end{eqnarray}
where $\delta \tilde g_{\mu\nu} = \tilde g_{\mu\nu} - \tilde g^{(0)}_{\mu\nu}$ 
with $\tilde g^{(0)}$ given by \eqref{eq:hatgdef} and 
$$
|\Omega^{-2} (\partial_x \Omega -1)|
+ \sum_{i=1}^3 |\Omega^{-3}\partial_i \Omega|<C 
$$
hold.
\item\label{point:Ricci}
The components of the Ricci tensor of the physical metric $g$, w.r.t. the 
coordinate frame are assumed to be bounded in $F({\cal N}\setminus \{x=0\})$.
\end{enumerate}
We will use the notation $\tilde V_E = F({\cal N}\cap \{u
>0\})$ and $V_E = \tilde V_E \cap V$ and similarly 
$V_I = V \setminus \text{int}(V_E)$. 
\end{definition}
If we demand that the metric $g$ satisfies Einstein's equation and that the 
stress energy tensor vanishes asymptotically to order 2 in $\Omega$,
Point \ref{point:Ricci} is immediately satisfied.
Note that $\Scri^+$ is given by 
$\{x=0 \}$ and that $\Scri^+$ is in the future of
$V$, i.e. it is future null infinity.
Further, $\tilde V_E$ is the subset of $F({\cal
N})$ which lies to the future of the  backward lightcone of the cut 
of $\Scri^+$ given by $\{ x=0, u=0\}$.

Some calculations will be done in the unphysical spacetime
$(\tilde V, \tilde g)$,
in which case   all the corresponding geometrical quantities will be denoted
with tilde and all the physical ones, i.e. calculated in $(V, g)$ will be
denoted without tilde. We remark that our convention for the usage of the
tilde
here coincides with that in \cite{HE} and is opposite of that in eg.
\cite{Geroch}.

\subsection{The reference foliation} \label{sec:ref-fol} 
The computations done in this Subsection will be used in Section 
\ref{sec:barriers} to construct a refence foliation which gives barriers 
near $\Scri^+$ in asymptotically Schwarzschild spacetimes.

Let $H$ be a $C^1$ function on $(0,x_0)\times (0,\infty)$ which extends as a
continuous function to $[0,x_0)\times (0,\infty)$. 
Further let $(V,g,{\cal N})$ and $(u,x,y^A)$ be as in Definition 
\ref{def:asymp-schwarz}.
Consider the one parameter family of embeddings 
$$
\Phi_{\tau} : (0,x_0) \times S^2 \to {\cal N}, \quad \tau \in (0,\infty) ,
$$
defined by 
$$
\Phi_{\tau}:  (x ,y^2,y^3)\to (H(x ,\tau),x ,y^2,y^3) 
$$
and let $S_{\tau} = \Phi_{\tau}((0,x_0) \times S^2)$. 
Then $S_\tau$ is the graph of a height function $H(\cdot ,\tau)$. 

The pushforward of the canonical frame to $S_{\tau}$ is
\begin{equation}
\label{eq:canonical}
\begin{array}{ll}
v_1 &=\Phi_* (\partial_x  )=\partial_x  H\partial_u+\partial_x ,\\
v_A &= \Phi_*(\partial_A )=\partial_A .
\end{array}
\end{equation}

In terms of this frame, the induced metric $\gamma$ on $S_{\tau}$ has 
components
\begin{equation}
\label{eq:gamind}
\begin{array}{rl}
\gamma_{11}&=(\partial_x  H)^2 g_{uu}+2\partial_x  H g_{ux} +g_{xx}\\
\gamma_{1A}&= \partial_x  H g_{uA}+g_{xA}\\
\gamma_{AB}&= g_{AB}.
\end{array}
\end{equation}

Assuming that $H$ is such that the foliation $\{S_{\tau}\}$ consists of spacelike
hypersurfaces, the future directed unit normal of $S_{\tau}$ is 
\begin{equation}
\hat T=-\eta (g^{u\nu}-g^{x\nu}\partial_x H )\partial_\nu,
\label{eq:hatT}
\end{equation}
where
\begin{equation}
\eta^{-2} =-\bigg (g^{uu}-2g^{xu}\partial_x H+g^{xx}(\partial_x H)^2\bigg ).
\label{eq:eta}
\end{equation}

Then the lapse and shift is given by 
\begin{equation}
\alpha =\eta\partial_{\tau} H \label{eq:alpha} 
\end{equation}
\begin{equation}\label{eq:beta}
\begin{array}{rl}
\beta_1&=\partial_\tau H(g_{uu}\partial_x  H + g_{ux})  \\
\beta_A&=\partial_\tau H g_{uA}, 
\end{array}
\end{equation}
where we have used $\gamma_{ij}$ to lower the index on $\beta^i$.

Let $k_{ij}= \langle v_i,\nabla_{ v_j} \hat T\rangle$ be the 
components of the second fundamental form of $S_{\tau}$ in the frame
$\{v_i \}$.
Then
\begin{equation}\label{eq:kijg}
\begin{array}{rl}
k_{11} &=\eta\big{(} (\partial_x  H)^2 \Gamma ^\mu_{uu} T_\mu
+ 2\partial_x  H\Gamma^\mu_{ux}T_\mu+\\
&\qquad \qquad \qquad \qquad +\Gamma^\mu_{xx} T_\mu
+\partial^2_x H \big{)} ,
 \\
k_{1A}&=\eta  ( \partial_x  H \Gamma^\mu_{uA}+\Gamma^\mu_{xA}) T_\mu ,
\\
 k_{AB} &=\eta \Gamma^\mu_{AB}T_\mu .
\end{array}
\end{equation}

Given the foliation  $\{ S_{\tau} \}$, we define a foliation 
$\{\tilde S_{\tau} \}$ of ${\cal N}$ by 
$\tilde S_{\tau} = S_{\tau} \cup \{ u=H(0,\tau) \}.$  
The second fundamental form of $\tilde S_{\tau}$ w.r.t. the unphysical metric
$\tilde g$ is 
\begin{equation}
\tilde k_{ij} =\Omega (k_{ij}+\lambda \gamma_{ij}),
\label{eq:kconf}
\end{equation}
where 
$\lambda= \tilde T(\Omega)= 
\tilde T^\alpha\frac{\partial\Omega}{\partial x^\alpha}$.
Here $\tilde T = \Omega^{-1} \hat T$ 
is the unit normal of $\tilde S_{\tau}$ w.r.t. $\tilde g$.
For a more explicit derivation  of these formulae see 
\cite[Section 3]{iriondo}.

\section{Barriers near $\Scri^+$}
\label{sec:barriers}
The goal of this section is to construct a foliation $\{S_{\tau}\}$
near $\Scri^+$ which 
can be used to get local barriers for the mean curvature operator 
near $\Scri^+$ in asymptotically Schwarzschild spacetimes.
This will allow us in Section
\ref{sec:existence} to construct a global upper barrier which is of central
importance in the proof of the existence result. Using $\{S_{\tau}\}$ as 
local lower barriers near $\Scri^+$ is essential in proving the 
completeness of CMC the hypersurface constructed in the existence part of
the argument.

In order to be able to use the $\{S_{\tau}\}$ as barriers we must check that
for sufficiently large $\tau$, the mean curvature $k$ of $S_{\tau}$ satisfies 
$k < 3$ and that for sufficiently small $\tau >0$,  $k> 3$.
Further to be able to apply the gradient estimates \cite[Theorem 3.1]{Bartnik}
we must check that the foliation $\{S_{\tau}\}$ satisfies the following
properties in a subset of $V_E$.
\begin{enumerate}
\item
Choose $T_1=\hat T$, where $\hat T$ is the normal of the foliation, then 
\begin{equation}\label{eq:max}
\max(\Vert \alpha\Vert ,\Vert
\alpha^{-1}\nabla\alpha\Vert, \Vert k(\cdot,\cdot)\Vert ,\Vert \hat T\Vert_2 )<C, 
\end{equation}
where the  norm  is taken  with the Euclidean metric $g_E$ given by
\eqref{eq:gEdef}.
\item
The Ricci tensor of the spacetime is  bounded, i.e.
\begin{equation}\label{eq:Ricci}
\Vert Ric\Vert<C.
\end{equation}
\end{enumerate}

\subsection{A foliation of the Schwarzschild spacetime}
\label{sec:fol-schwarz}
Consider the Schwarzschild metric in the $(t,r,\theta,\phi)$ coordinate
system. Let $A=1-2m/r$. 
The height function
$$
w(\tau, r) =\sqrt{\tau^2+(r-2m)^2}+2m\log\big (r-2m+\sqrt{\tau^2+(r-2m)^2}\big )- 2m (\log 2-1),
$$
is the solution to the radial mean curvature equation in the 
Schwarzschild  spacetime, 
\begin{eqnarray}
k(w)&=&\bigg (\frac{w'Ar^2}{\sqrt{1-(Aw^{\prime2})}}\bigg )'r^{-2}\sqrt{A}\nonumber\\
&=&\frac{3}{\tau}\sqrt{A}\bigg(1-\frac{4m}{3r}\bigg),
\label{eq:diffw}
\end{eqnarray}
satisfying
$$
\lim_{\tau \to 0} w(\tau,r) =  r+2m\log(r-2m),
$$
i.e. the level set $t =w(\tau,r)$ tends to the light cone as $\tau \to 0$, cf 
\cite[Section 3]{iriondo}.

Let $\{u,x ,y^A\}_{A=2,3}$ be a coordinate system for the Schwarzschild
spacetime, where $u$ is the retarded coordinate and $x=1/r$ is the conformal 
factor. In this coordinate system the foliation given
by $t=w(\tau,r)$ is given by $u = H(\tau,x)$ where 
\begin{eqnarray}
H(x,\tau) &=&w(\tau,1/x)-\frac{1}{x}-2m\log\bigg(\frac{1}{x}-2m\bigg)\nonumber\\
&=&
\frac{1}{x }\big (\sqrt{\tau^2x ^2+h^2(x )}-h(x )\big )\nonumber\\
& &+2m\big (\log(h(x )+\sqrt{\tau^2x ^2+h^2(x )})-\log h(x )-\log 2\big),
\label{eq:height-u-Om1}
\end{eqnarray}
where $h(x)$ is given by \eqref{eq:h(x)def}. The partial derivatives of the height function $H$ are 
\begin{equation}
\partial_x H=-\frac{1}{x^2}\bigg (\frac{1}{\sqrt{\tau^2 x ^2+h^2(x)}}-
\frac{1}{h
(x)}\bigg )
\label{eq:dH}
\end{equation}
and
\begin{equation}
\partial_{\tau} H= \frac{\tau x }{\sqrt{\tau^2 x ^2+h^2(x)}}\bigg
(1+\frac{2mx }{
h(x )+\sqrt{\tau^2x ^2+h^2(x)}}\bigg ) .
\label{eq:dfntau}
\end{equation}

In the two following lemmata  we  summarize  the properties of  the foliation  
given by the height function \eqref{eq:height-u-Om1} in the Schwarzschild spacetime. The calculations 
are straightforward and  they are done in detail in \cite[Section 3]{iriondo}. 
The function given by \eqref{eq:height-u-Om1} will  be used in the 
construction of the barriers in asymptotically  Schwarzschild spacetimes.
\begin{lemma}
\label{lem:fol1}
Let $(V,g)$ be the Schwarzschild spacetime and let $\tilde V_E$ be as in
Definition \ref{def:asymp-schwarz}. There is a foliation 
$\{ \tilde S_{\tau}\}_{\tau \in (0,\infty)}$ of $\tilde V_E$ and a foliation
$\{S_{\tau}\}_{\tau \in (0,\infty)}$ of $V_E$, where  
$S_{\tau} = \tilde S_{\tau} \cap V$, 
such that the following is true.
\begin{enumerate}
\item
The induced metric $\gamma$ is Riemannian and conformally related to a metric
which is  $C^{\infty}$ on $\tilde S_{\tau}$.
\item
For each $\tau$, $\tilde S_{\tau} \cap \Scri ^+ \approx S^2$.
\item
The lapse function satisfies  $\alpha \big{|}_{\Scri^+} =1$.
\item
The shift vector satisfies $\beta \big{|}_{\Scri^+}=0$.
\item
$|\hat T(\alpha)|\leq \frac{Cx}{\tau}$ and 
$|\hat T(v_1(\alpha))|<\frac{C}{\tau}$.
\end{enumerate}
\end{lemma}
\begin{pf}
Using \eqref{eq:gamind} we get 
\begin{equation}
 \Phi^*_\tau(\tilde g^{(0)})=\frac{\tau^2}{h(x )(\tau^2x ^2+h^2(x ))}dx ^2 +d\theta^2+\sin^2\theta d\phi^2.
\label{eq:hatgam}
\end{equation}
Further, using \eqref{eq:alpha} and \eqref{eq:beta} we have 
\begin{eqnarray}
\alpha&=& \sqrt{h(x )}\bigg(1+\frac{2mx }{h(x )+\sqrt{\tau^2x ^2+h^2(x )}}\bigg),
\label{eq:schwalpha}\\ 
\beta&=&\frac{h^2(x )x }{\tau }\bigg(1+\frac{2mx }{h(x )+\sqrt{\tau^2x ^2+h^2(x )}}\bigg)v_1 \nonumber \\
&=&\frac{(h(x))^{\frac{3}{2}}\alpha x}{\tau} v_1. \label{eq:schwbeta}
\end{eqnarray}
Using the lapse and the shift we write
\begin{eqnarray*}
\hat T(\alpha)&=& \alpha^{-1}\big (\partial_\tau\alpha-\beta(\alpha)\big )\\
\hat T(v_1(\alpha))&=&\alpha^{-1}\big (\partial_\tau\partial_x\alpha-\beta(\partial_x\alpha)\big )
\end{eqnarray*}
and a   straightforward calculation shows that
\begin{equation}
|\partial_\tau\alpha|<\frac{Cx}{\tau},\quad |\partial_x  \alpha| <C,\quad|\partial_\tau\partial_x\alpha|<\frac{C}{\tau}\quad\text{and}\quad
|\partial^2_x  \alpha| <C,
\label{eq:dalpha}
\end{equation}
then
$$
|\hat T(\alpha)|<\frac{Cx}{\tau} \qquad\text{and}\qquad |\hat T(v_1(\alpha))|<\frac{C}{\tau}.
$$
\end{pf}
Beside the properties given by this Lemma the foliation defined by \eqref{eq:height-u-Om1} has 
a second fundamental form with similar behavior to that of the 
hyperboloids in  Minkowski space. 
\begin{lemma}[\protect{\cite[Lemma 3.2]{iriondo}}]\label{lem:schwlemma}
Let $(V,g)$ and $\{\tilde S_{\tau}\}, \{ S_{\tau} \}$ be as in Lemma \ref{lem:fol1}.
The physical and unphysical second fundmamental forms $k_{ij}$ and 
$\tilde k_{ij}$ of $S_\tau$ and $\tilde S_{\tau}$, respectively in the canonical frame given by \eqref{eq:canonical}, are
of the form
$$
k_{ij}=\frac{1}{\tau}\left (\gamma_{ij}+ r_{ij}\right )
$$
where $|r_{ij}| ={\cal O}(x )$ and
$$
\tilde k_{11}=\frac{2m\sqrt{h(x)}}{\tau}\tilde \gamma_{11}\qquad 
\qquad\tilde
k_{ij}= 0 \text{ for } (i,j) \ne (1,1).
$$
\end{lemma}
\subsection{A foliation of asymptotically Schwarzschild spacetimes}
\label{sec:fol-asympscwarz}
Next we will see that the foliation introduced in Subsection \ref{sec:fol-schwarz}
gives local  barriers for the mean curvature equation in asymptotically
Schwarzschild spacetimes. 
Let $0 < \tau_0^- < \tau_0^+ < \infty$ be given. 
\begin{lemma} \label{lem:Hprop}
The function given by \eqref{eq:height-u-Om1} has  the following properties:
\begin{enumerate}
\item
For each fixed $x$
$$
\lim_{\tau\to\infty}\partial_x H=\frac{1}{x^2h(x)}.
$$
\item
For all $\tau>\tau_0^+$ 
$$
\partial_xH={\cal O}(x^{-2}),\quad \partial ^2_xH=x^{-1}\partial_x H
+{\cal O}(x^{-2}).
$$
\item
For all $\tau<\tau_0^-$ 
$$
\partial_x H=\tfrac{1}{2}\tau^2(1+{\cal O}(x)).
$$
\end{enumerate}
\end{lemma}

\mnote{\lars I commented out a remark here, do you want to put something 
in to clarify??}
\mnote{I dont understand what kind of comment, I have change $\epsilon$ with $\epsilon_-$}

Let $\epsilon_- > 0$ be given.
We denote by $V_E^+$ and $V_E^-$ be given by
\begin{equation}\label{eq:ve}
\begin{array}{rl}
V_E^+&=V_E\cap \{\cup_{\tau\geq\tau_0^+}S_{\tau}\},
\\
V_E^-&=V_E\cap \{\cup_{\epsilon<\tau\leq\tau_0^-}S_{\tau}\},
\end{array}
\end{equation}

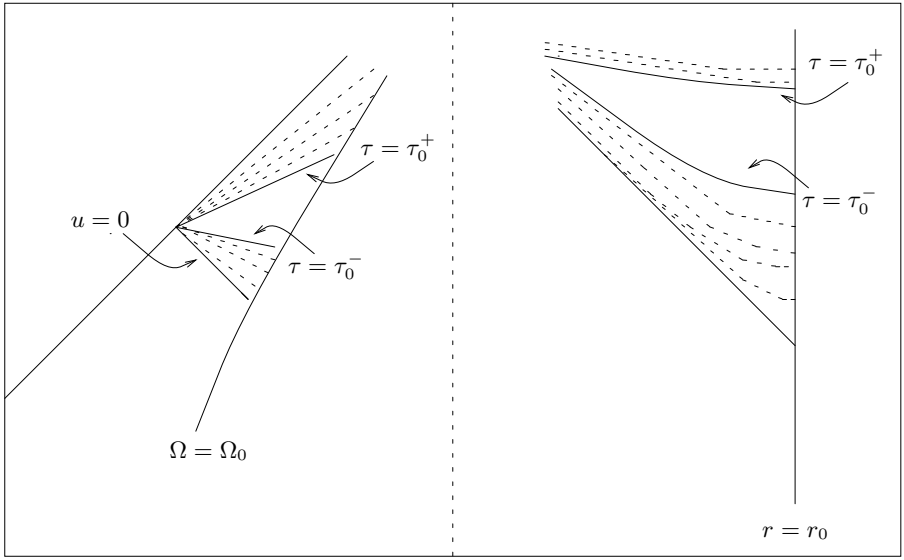
\begin{figure}
\centering
\setlength{\unitlength}{0.0069in}
\begin{picture}(680,435)(0,-10)
\path(260,380)(0,120)
\path(130,250)(185,195)(180,200)
\drawline(80,245)(80,245)
\drawline(420,380)(420,380)
\dashline{4.000}(340,420)(340,0)
\path(680,0)(680,420)(0,420)
        (0,0)(680,0)
\path(420,340)(600,160)
\path(600,400)(600,40)
\dashline{4.000}(130,250)(265,325)
\dashline{4.000}(130,250)(280,350)
\dashline{4.000}(130,250)(280,370)
\dashline{4.000}(135,245)(190,205)
\dashline{4.000}(135,250)(200,215)
\dashline{4.000}(135,250)(205,225)
\dashline{4.000}(410,390)(545,370)(600,370)
\dashline{4.000}(410,385)(570,360)(595,360)
\dashline{4.000}(415,365)(550,260)(600,250)
\dashline{4.000}(420,355)(485,295)(535,250)
\dashline{4.000}(535,250)(570,235)(600,230)
\dashline{4.000}(420,345)(490,275)(520,250)
        (560,225)(585,220)(595,220)
\dashline{4.000}(455,305)(560,210)(590,195)(600,195)
\path(130,250)(205,235)
\path(130,250)(250,305)
\spline(90,245)
(125,215)(145,225)
\path(138.739,219.633)(145.000,225.000)(136.950,223.211)
\spline(280,300)
(250,280)(230,290)
\path(238.050,288.211)(230.000,290.000)(236.261,284.633)
\spline(235,230)
(205,255)(190,245)
\path(195.547,251.102)(190.000,245.000)(197.766,247.774)
\spline(415,370)
(530,285)(600,275)
\spline(610,280)
(580,305)(565,295)
\path(570.547,301.102)(565.000,295.000)(572.766,297.774)
\spline(640,360)
(610,340)(590,350)
\path(598.050,348.211)(590.000,350.000)(596.261,344.633)
\spline(410,380)
(520,360)(600,355)
\spline(290,365)
(180,185)(145,95)
\put(50,250){\makebox(0,0)[lb]{\raisebox{0pt}[0pt][0pt]{\shortstack[l]{{\footnotesize $u=0$}}}}}
\put(215,215){\makebox(0,0)[lb]{\raisebox{0pt}[0pt][0pt]{\shortstack[l]{{\footnotesize $\tau=\tau^-_0$}}}}}
\put(125,75){\makebox(0,0)[lb]{\raisebox{0pt}[0pt][0pt]{\shortstack[l]{{\footnotesize $\Omega=\Omega_0$}}}}}
\put(575,15){\makebox(0,0)[lb]{\raisebox{0pt}[0pt][0pt]{\shortstack[l]{{\footnotesize $r=r_0$}}}}}
\put(270,305){\makebox(0,0)[lb]{\raisebox{0pt}[0pt][0pt]{\shortstack[l]{{\footnotesize $\tau=\tau^+_0$}}}}}
\put(610,370){\makebox(0,0)[lb]{\raisebox{0pt}[0pt][0pt]{\shortstack[l]{{\footnotesize $\tau=\tau^+_0$}}}}}
\put(605,265){\makebox(0,0)[lb]{\raisebox{0pt}[0pt][0pt]{\shortstack[l]{{\footnotesize $\tau=\tau^-_0$}}}}}
\end{picture}
\caption[]{The unphysical  and the physical picture ($r = 1/x$)}
\label{fig:fig6}
\end{figure}

We claim that in the region given by $V_E^+\cup V_E^-$, (see Figure \ref{fig:fig6}),
the foliation $\{S_{\tau} \}$ behaves similarly in an 
asymptotically Schwarzschild spacetime as in the standard Schwarzschild
spacetime. 

\begin{lemma}
\label{lem:fol1b}
Let $(V,g)$ be an asymptotically  Schwarzschild spacetime and let $V_E^+$ be
given by \eqref{eq:ve}. 
There is a foliation $\{ \tilde S_{\tau}\}_{\tau \in (\epsilon_-,\infty)}$ of 
$\tilde V_E^+$ and a foliation $\{S_{\tau}\}_{\tau \in (\epsilon_-,\infty)}$ of\mnote{introduce $\epsilon_-$ as in the definition above} 
$V_E^+$, where $S_{\tau} = \tilde S_{\tau} \cap V$,
such that the following is true.
\begin{enumerate}
\item \label{point:gam1}
The induced metric $\gamma$ is Riemannian and of the form 
$\gamma_{ij}=x^{-2}( \tilde \gamma^{(0)}_{ij}+ r_{ij})$, where 
where $\tilde \gamma^{(0)}_{ij}$ is the metric given by \eqref{eq:hatgam} and 
$|r_{ij}| \leq C x^2$ and 
\item
$\tilde S_{\tau} \cap \Scri ^+ \approx S^2$.
\item \label{point:alpha}
The lapse function satisfies  $\alpha \big{|}_{\Scri^+} =1$.
\item \label{point:shift}
The shift vector satisfies $\beta \big{|}_{\Scri^+}=0$ and 
$|v_j(\beta^i)|<C$.
\item \label{point:T}
$ |\hat T(\alpha)|\leq \frac{Cx}{\tau}+Cx^3$ and $\max (\Vert \alpha^{-1}\nabla \alpha\Vert,|\hat T(v_i(\alpha))|) < C$ .
\end{enumerate}
The constants depend only on the choice of $\tau_0^+$ and $x_0$, and the frame used on $\tilde S_\tau$ and $S_\tau$ is the canonical one given by \eqref{eq:canonical}.
\end{lemma}
\begin{pf}
As in Lemma \ref{lem:fol1} we compute
\begin{eqnarray}
\tilde \gamma_{11}&=&-x^2h(x)(\partial_xH)^2 +2(\partial_x   H)+(\partial_xH)^2 \delta  \tilde g_{uu}\nonumber\\
&=&\frac{\tau^2}{(\tau^2 x ^2+h^2(x ))h(x)}+\delta  \tilde g_{uu}(\partial_xH)^2,
\label{eq:tilde11+}
\end{eqnarray}
from \eqref{eq:dH} we  calculate
\begin{equation}
(\partial_xH)^2=\frac{\tau^2\epsilon_0(x,\tau)}{x^4h(x)(\tau^2 x ^2+h^2(x ))},
\label{eq:dH2}
\end{equation}
where
\begin{equation}
\epsilon_0(x,\tau)=\frac{1}{h(x)}\bigg (x^2+2(\frac{h(x)}{\tau})^2-2\frac{h(x)}{\tau}\sqrt{x ^2+(\frac{h(x )}{\tau})^2}\bigg ),
\label{eq:epsilon0}
\end{equation}
For each $\tau$, $\epsilon_0\big{|}_{\Scri^+}=0$. 
Now \eqref{eq:tilde11+} can be written in the form
\begin{equation}
\tilde \gamma_{11}=\frac{\tau^2(1+\epsilon_1)}{(\tau^2 x ^2+h^2(x ))h(x)},
\label{eq:gam1}
\end{equation}
where
\begin{equation}
\epsilon_1(x^\mu,\tau)= \epsilon_0(x,\tau) \delta \tilde g_{uu}x^{-4}.
\label{eq:eps1}
\end{equation}
It follows from \eqref{eq:asest1} that $\epsilon_1 = {\cal O}(x^2)$.

A calculation shows that the other components of the metric satisfy
\begin{eqnarray}
\tilde \gamma_{1A} &= &{\cal O}(x)\nonumber\\
\tilde \gamma_{AB}&=&\tilde g_{AB},
\label{eq:gam1AB}
\end{eqnarray}
and Point \ref{point:gam1} follows.

It follows from Definition \ref{def:asymp-schwarz} that  
the inverse of metric takes the form
\begin{equation}\label{eq:invmetr}
(\tilde g^{\mu\nu})=\left(\begin{array}{cccc}
    0 &  1     & 0         & 0\\
1&x^2h(x)+{\cal O}(x^6)  \;   & {\cal O}(x^3) &{\cal O}(x^3)\\
0&  {\cal O}(x^3) &     & \\
0&  {\cal O}(x^3) &
\multicolumn{2}{c}{\smash{\raise6pt\hbox{$\tilde g^{AB}$}}}
         \end{array}     \right)  .
\end{equation}
Let $\delta \tilde g^{\mu\nu} = \tilde g^{\mu\nu} - \tilde g^{(0)\,\mu\nu}$.
Using equation \eqref{eq:eta} we have
$$
\tilde \eta^{-2} =\frac{\tau^2}{(\tau^2 x ^2+h^2(x ))h(x)}+\delta \tilde
g^{xx}(\partial_xH)^2 .
$$
Letting
$$
\epsilon_2(x^\mu,\tau)= \epsilon_0(x,\tau) \delta \tilde g^{xx}x^{-4} ,
$$
we get
$$
\tilde \eta^{-2}=\frac{\tau^2(1+\epsilon_2)}{(\tau^2 x ^2+h^2(x ))h(x)}.
$$
By Equation \eqref{eq:invmetr}, $\epsilon_2 = {\cal O}(x^2)$.
In terms of the physical metric, $\eta$ becomes
\begin{eqnarray*}
\eta&=&\frac{\sqrt{(\tau^2 x ^2+h^2(x ))h(x)}}{\tau x\sqrt{1+\epsilon_2}}(1+{\cal O}(x^2))\\
&=&\frac{\eta^{(0)}}{\sqrt{(1+\epsilon_2)}}(1+{\cal O}(x^2)) ,
\end{eqnarray*}
where $\eta^{(0)}$ is the quantity corresponding to $\eta$ in the
Schwarzschild metric.

Using \eqref{eq:dfntau} and \eqref{eq:alpha} we  get
$$
\alpha= \frac{\alpha^{(0)}}{\sqrt{1+\epsilon_2}}(1+{\cal O}(x^2)),
$$
where $\alpha^{(0)}$ is given by \eqref{eq:schwalpha}. This proves Point 
\ref{point:alpha}.
Further, using  \eqref{eq:beta} we get
\begin{eqnarray*}
 \beta_1 &=&\beta_1^{(0)}+{\cal O}(x^2)\\
\beta_A&=&\partial_\tau H g_{uA} ={\cal O}(x),
\end{eqnarray*}
where $\beta_i^{(0)}$ denotes the components of the dual w.r.t. the metric 
$x^{-2} \tilde \gamma^{(0)}$ of the vector given by \eqref{eq:schwbeta}.
Thus raising the index with the metric $\gamma$ we get $\lim_{x\to 0}\beta = 0$.
Now we will estimate $v_i(\beta^j)$.
Observing that 
$\beta^i = \tilde\beta^i$,
we find
$$
v_l(\tilde \beta^i)=\tilde \gamma^{ij}v_l(\tilde \beta_j)+v_l(\tilde \gamma^{ij})\tilde \beta_j.
$$
The most delicate calculation is  $v_1(\tilde \beta_1)$ and 
$v_1(\tilde \gamma^{11})$, due to the fact that the height function $H$ is
independent of the coordinates $(y^A)$. 

By \eqref{eq:gam1} we can estimate these derivatives if we can estimate 
$v_1(\epsilon_1)$. 
Note that $v_1=\partial_xH\partial_u+\partial_x$ and thus, if 
$f=f(\tau,x)$ then $v_1(f)=\partial_xf$ because $v_1(\tau)=0$.
Therefore 

\begin{equation}
v_1( \epsilon_1)= \partial_x\epsilon_0(x,\tau)  \delta \tilde g_{uu}x^{-4}+\epsilon_0(x,\tau) v_1(\delta \tilde g_{uu})x^{-4}-4\epsilon_0(x,\tau) \delta \tilde g_{uu}x^{-5}.
\label{eq:deps-1}
\end{equation}
A straightforward calculation shows that
\begin{equation}
|\partial_x\epsilon_0|<C.
\label{eq:v1eps0}
\end{equation}
Using this and the decay of the metric we obtain 
$|v_j(\beta^i)|<C$. This proves Point \ref{point:shift}.

It remains to prove Point \ref{point:T}.
Observe that the normal of the foliation can be written using different 
frames as follows
\begin{eqnarray}
\hat T&=&-\Omega \tilde \eta\bigg(-\partial_xH\partial_u
+\big(\frac{h(x)}{\sqrt{h^2(x)+x^2\tau^2}}
+{\cal O}(x^4)\big )\partial_x+ {\cal O}(x)\partial_A\bigg)\label{eq:normalu}\\
&=&\alpha^{-1}(\partial_\tau-\beta^iv_i).
\label{eq:normal}
\end{eqnarray}
If $\hat T$ acts on a function $f=f(\tau,x)$ we use \eqref{eq:normal} while
if $\hat T$ acts on a function $f= f(u,x)$ we use \eqref{eq:normalu}.
Note that 
$$
\alpha=\frac{\alpha^{(0)} x^{-1}\Omega}{\sqrt{1+\epsilon_2}}.
$$
Therefore
$$
\hat T(\alpha)=\hat T(\alpha^{(0)})
\frac{  x^{-1}\Omega}{\sqrt{1+\epsilon_2}}
-\frac{\alpha}{2(1+\epsilon_2)} 
\hat T(\epsilon_2)+\frac{\alpha^{(0)}}{\sqrt{1+\epsilon_2}}
\hat T(\Omega x^{-1})
$$
and
\begin{equation}
\hat T(\epsilon_2)=\hat T(\epsilon_0)\delta \tilde g^{xx}x^{-4}
+\epsilon_0(x,\tau) \hat T(\delta \tilde g^{xx})x^{-4}
-4\epsilon_0(x,\tau) \delta \tilde g^{xx}\hat T(x^{-4}).
\label{eq:Teps-2}
\end{equation}
A straightforward calculation gives
\begin{equation}
|\partial_\tau\epsilon_0|<\frac{Cx^3}{\tau^3}.
\label{eq:taueps0}
\end{equation}
Using \eqref{eq:taueps0} and \eqref{eq:v1eps0} we estimate
\begin{eqnarray*}
\hat T(\epsilon_0)&=&\alpha^{-1}(\partial_\tau\epsilon_0-\beta(\epsilon_0))\\
&=&\frac{{\cal O}(x)}{\tau}.
\end{eqnarray*}
On the other hand
\begin{eqnarray*}
\Omega\hat T(x^{-1})&=&-(x^{-1}\Omega)^2 \tilde T(x)\\
&=&(x^{-1}\Omega)^2 \tilde \eta \big(\frac{h(x)}{\sqrt{h^2(x)+x^2\tau^2}}
+{\cal O}(x^4)\big )
\end{eqnarray*}
and using \eqref{eq:normal} and the decay of the derivatives of the conformal factor, 
we get the estimate
\begin{equation}
x^{-1}\hat T(\Omega )=
-x^{-1}\Omega\tilde \eta \bigg
(\frac{h(x)}{\sqrt{h^2(x)+x^2\tau^2}}+{\cal O}
(x)\bigg).
\label{eq:TOmega}
\end{equation}
Thus 
$$
|\hat T(x^{-1}\Omega)|<\frac{Cx}{\tau}+Cx^3
$$
and hence
\begin{equation}
|\hat T(\alpha)|<\frac{Cx}{\tau}+Cx^3.
\label{eq:Talpha}
\end{equation}
Let  $\{\hat T,\hat v_i \}$ be the frame given by  the normal  of the 
foliation and  $\hat v_i=v_i/\sqrt{\langle v_i, v_i\rangle}$. Then
\begin{eqnarray*}
\Vert \alpha^{-1}\nabla \alpha\Vert^2 &=& 
\alpha^{-2} g_E(\nabla \alpha,\nabla \alpha) \\
&=&\alpha^{-2} \bigg ((\hat T(\alpha))^2+\sum _{i,j=1}^3
\frac{v_i (\alpha) v_j (\alpha)\langle v_i, v_j\rangle}
{\sqrt{\langle v_i, v_i\rangle\langle v_j, v_j\rangle}}\bigg ).
\end{eqnarray*}
Therefore we need to estimate 
$v_i(\alpha)$. The delicate step is to get a bound for 
$$
v_1(\epsilon_2)=\partial_x\epsilon_0(x,\tau)  \delta \tilde g^{xx}x^{-4}
+\epsilon_0(x,\tau) v_1(\delta \tilde g^{xx})x^{-4}
-4\epsilon_0(x,\tau) \delta \tilde g^{xx}x^{-5}.
$$
Using \eqref{eq:v1eps0}, \eqref{eq:Talpha} and the decay of the metric gives the
desired estimate.

Finally, a straightforward calculation shows that
$$
|\partial_\tau\partial_x\epsilon_0|<C
\qquad\text{and}\qquad|\partial_x^2\epsilon_0|<C.
$$
Using this, the decay of the metric  and Lemma \ref{lem:fol1},  
we obtain that $|\hat T(v_i(\alpha))|$ is bounded.
\end{pf}

In order to compute the second fundamental form of $S_{\tau}$ we will first do 
the calculations in the 
unphysical spacetime and then using the conformal transformation given by 
\eqref{eq:kconf}, we obtain 
$k_{ij}$.
\begin{lemma}
\label{lem:kt}
Let $(V,g)$ and $\{\tilde S_{\tau}\}, \{ S_{\tau} \}$ be as in Lemma
\ref{lem:fol1b}.
The unphysical and physical second fundmamental forms $\tilde k_{ij}$ and
$k_{ij}$ of $\tilde S_\tau$ and $S_{\tau}$, respectively, satisfy the
estimates
\begin{equation}\label{eq:tildektest}
|\tilde k_{11}|\leq C\tilde\gamma_{11}\bigg(\frac{1}{\tau}+x^2\bigg),
\quad |\tilde k_{1A}|\leq C\bigg(\frac{1}{\tau}+x\bigg),\quad 
|\tilde k_{AB}|\leq C\bigg(\frac{1}{\tau}+x\bigg) ,
\end{equation}
where $\tilde k_{ij}$ is computed in the coordinate frame on 
$T\tilde S_\tau$ and $C$ is some constant depending on $\tau_0^+$ and $x_0$,
and 
\begin{equation}\label{eq:ktest}
|k(\hat v_i,\hat v_j)|\leq  \frac{1+\epsilon}{\tau}+C x^2,
\end{equation}
where $ \hat v_i= v_i/\sqrt{\langle v_i, v_i\rangle}$, and $\epsilon$ and $C$
are constants depending on $\tau_0^+$ and $x_0$.
\end{lemma}
\begin{pf}
Using the corresponding formula to \eqref{eq:kijg} for the unphysical metric,  we calculate the components
of $\tilde k_{ij}$. 
The most complicated calculation will be in the $\{v_1,v_1\}$ direction. 

By  \eqref{eq:delga1} we get that $\tilde \Gamma ^u_{xx}$, $\tilde \Gamma ^x_{xx}$ and $\tilde \Gamma ^u_{xu}$ are zero, hence 
we obtain
\begin{equation}
\tilde k_{11}=\tilde k^{(0)}_{11} +\tilde \eta(\partial_x  H)^2 (\delta \tilde  \Gamma ^u_{uu}- \partial_x  H\delta\tilde\Gamma ^x_{uu}- 2\delta\tilde \Gamma^x_{ux} ),
\label{eq:tk11}
\end{equation}
where $\tilde k^{(0)}_{ij}$ is the second fundamental form for $\tilde S_{\tau}$
in the unphysical Schwarzschild spacetime given in Lemma
\ref{lem:schwlemma} and 
$\delta\tilde \Gamma=\tilde  \Gamma-\tilde \Gamma^{(0)}$. 
Using \eqref{eq:dH2} and Lemma \ref{lem:fol1b}, we  write
$$
(\partial_xH)^2=\frac{\tau^2\epsilon_0(x,\tau)}{x^4h(x)(\tau^2 x ^2+h^2(x ))}
\quad\text{and}\quad 
\tilde \eta=\frac{\sqrt{(\tau^2 x ^2+h^2(x ))h(x)}}{\tau \sqrt{1+\epsilon_2}},
$$
hence in order to estimate the second term in \eqref{eq:tk11} we use the triangle
inequality to get 
$$
\epsilon_0\sqrt{x ^2+\bigg(\frac{h(x)}{\tau}\bigg)^2}
\leq \frac{x}{\tau}\bigg(\frac{2}{\tau}+{\cal O}(x)\bigg)+x^3(1+{\cal O}(x)).
$$
Now we obtain
\begin{equation}
(\partial_xH)^2\tilde \eta<\frac{\tilde\gamma_{11}}{x^3\tau(1+\epsilon_1)\sqrt{1+\epsilon_2}}\bigg (\frac{2}{\tau}+
{\cal O}(x)\bigg )+\frac{\tilde\gamma_{11}}{x(1+\epsilon_1)\sqrt{1+\epsilon_2}}
\label{eq:dH2eps}
\end{equation}
and  using \eqref{eq:delga1} we estimate
\begin{equation}
|\delta \tilde  \Gamma ^u_{uu}- \partial_x  H\delta\tilde\Gamma ^x_{uu}- 2\delta\tilde \Gamma^x_{ux}|<Cx^3.
\label{eq:est-Gam}
\end{equation}

Inserting  \eqref{eq:est-Gam} and \eqref{eq:dH2eps} in the second term of \eqref{eq:tk11}, we have
\begin{eqnarray}
&& |\tilde \eta(\partial_x  H)^2 (\delta \tilde  \Gamma ^u_{uu}
- \partial_x  H\delta\tilde\Gamma ^x_{uu}
- 2\delta\tilde \Gamma^x_{ux} )|\nonumber\\
&& \hskip 0.9in <\frac{\tilde\gamma_{11}}{\tau}
\left (\frac{2C}{\tau}+{\cal O}(x)\right )+Cx^2\tilde\gamma_{11}.
\label{eq:term2}
\end{eqnarray}
On the other hand, $\tilde k^{(0)}_{11}$ can be written in terms of $\tilde \gamma_{11}$ as

$$
\tilde k^{(0)} _{11} =\frac{\tilde\gamma_{11}}{\tau}\frac{2m\sqrt{h(x)}}{1+\epsilon_1},
$$
inserting  this and \eqref{eq:term2} in \eqref{eq:tk11} we get
$$
|\tilde k_{11}|\leq C\tilde\gamma_{11}(\frac{1}{\tau}+x^2).
$$
In the other directions we easily estimate
\begin{eqnarray*}
\tilde k_{1A}&=&\tilde \eta ( \partial_x  H\tilde  \Gamma^\mu_{uA}+\tilde \Gamma^\mu_{xA}) T_\mu\\
&=&\tilde \eta\partial_x  H (\partial_x  H\tilde \Gamma^x_{uA}+\tilde \Gamma^x_{xA}+\tilde\Gamma^u_{uA})\\
&=&\tilde \eta{\cal O}(1)\\
\tilde k_{AB}&=&\tilde \eta\tilde \Gamma^\mu_{AB} T_\mu,\\
&=&\tilde \eta{\cal O}(1),
\end{eqnarray*}
which proves \eqref{eq:tildektest}.

The relation between the physical and unphysical second fundamental forms 
after a conformal transformation is given by equation \eqref{eq:kconf}.
$\lambda$ can be estimated using  \eqref{eq:TOmega} by 
\begin{eqnarray*}
\lambda&=&\tilde T(\Omega)\\
&=&-\tilde \eta \bigg (\frac{h(x)}{\sqrt{h^2(x)+x^2\tau^2}}+{\cal
O}(x)\bigg)\\
&=&-\frac{1+{\cal O}(x)}{\tau\sqrt{1+\epsilon_2}}+{\cal O}(x^2) .
\end{eqnarray*}
Using the estimates \eqref{eq:tildektest} completes the proof.
\end{pf}
In order to obtain an apriori estimate of $\nu$ in $V_E^+$, we need to prove 
that $\Vert \hat T\Vert_2$ is uniformly bounded.
For this purpose we will calculate the components of the commutator 
$[\hat v_i,\hat T]$.

\begin{lemma}
\label{lem:comut-dk}
\mnote{state more precisely, write as an inequality!!, do you mean $\Vert[\hat v_i,\hat T]\Vert<C$?}
Let $(V,g)$ satisfy the conditions of the last Lemma.
Then the components of the Lie-brackets  $[\hat v_i,\hat T]$ are bounded in 
the frame $\{\hat T, \hat v_i\}$.
\end{lemma}
\begin{pf}
From 
$$
[v_i,\hat T]= -\alpha^{-1}v_i(\alpha)\hat T-\alpha^{-1}v_i(\beta^j)v_j,
$$
we get 
\begin{equation}
[\hat v_i,\hat T]=\frac{1}{\sqrt{\langle v_i,v_i\rangle}}[v_i,\hat T]+\frac{\hat T(\langle v_i,v_i\rangle)}{\langle v_i,v_i\rangle}\hat v_i.
\label{eq:comut}
\end{equation}
Computing the Lie derivative of the metric $\gamma$, we get
$$
\frac{\hat T(\langle v_i,v_i\rangle)}{\langle v_i,v_i\rangle}=2k(\hat v_i,\hat v_i)+\frac{2\gamma_{il}\nabla_{v_i}\hat T^l}{\langle v_i,v_i\rangle}
$$
and inserting this in \eqref{eq:comut} we obtain
\begin{eqnarray}
\langle[\hat v_i,\hat T],\hat T\rangle&=&\alpha^{-1}\hat v_j(\alpha) \nonumber\\
 \langle[\hat v_i,\hat T],\hat v_j\rangle&=&\bigg (v_i(\beta^j)+2k(\hat v_i,\hat v_i)+ \frac{2\gamma_{il}\nabla_i\hat T^l}{\langle v_i,v_i\rangle}\bigg )\langle \hat v_j,\hat v_k\rangle.
\label{eq:comutcomp}
\end{eqnarray}
The proof is completed  by  referring to Lemma \ref{lem:fol1b} and Lemma \ref{lem:kt}.
\end{pf}

\begin{lemma}
\label{lem:T2}
Let $(V,g)$ be an asymptotically Schwarzschild spacetime at null infinity, then the normal of the foliation given by \eqref{eq:height-u-Om1}, satisfies
$$
\Vert \hat T\Vert_2 <C,\quad \text{in }  V_E^+.
$$
\end{lemma}
\begin{pf}
We choose on $TS_\tau$ a ON-frame given by $z_i= A_{ij}\hat v_i$, where $A$
is the 
matrix which orthogonalizes $\hat v_i$. It should be pointed out that by
Lemma \ref{lem:fol1b}, in a first order approximation this matrix is the identity.
Because
 $\langle \hat T,\hat T\rangle=-1$, we can write
$$
\nabla_{z_i}\hat T =\langle z_j,\nabla_{z_i}\hat T\rangle z_j,
$$
and by 
\eqref{eq:comutcomp}
\begin{eqnarray*}
\nabla_{\hat T}\hat T &=&\langle z_j,\nabla_{\hat T}\hat T\rangle z_j\\
&=&-\langle \nabla_{\hat T}z_j,\hat T\rangle z_j\\
&=&\langle [z_i,\hat T],\hat T\rangle z_j\\
&=&\alpha^{-1}z_j(\alpha) z_j,
\end{eqnarray*}
hence using Lemma \ref{lem:fol1b} and Lemma \ref{lem:comut-dk}  we get 
\begin{equation}\label{eq:nablaTest}
g_E(\nabla \hat T,\nabla \hat T) <C.
\end{equation}
We see from the above that we only need to estimate 
$g_E(\nabla \nabla \hat T,\nabla \nabla \hat T)$. 
Let $\{\hat T,z_i \}$ be an ON-frame. We shall estimate the different 
components of the tensor $\nabla \nabla \hat T$, namely

\mnote{can we rewrite here so that we do not need to display all of this?? I dont know how to do here}
\begin{eqnarray*}
\langle \hat T,\nabla_{\hat T}\nabla_{\hat T}\hat T\rangle&=&-\langle \nabla_{\hat T}\hat T,\nabla_{\hat T}\hat T\rangle\\
\langle z_i,\nabla_{\hat T}\nabla_{\hat T}\hat T\rangle&=&\nabla_{\hat T}\langle z_i,\nabla_{\hat T}\hat T\rangle-
\langle \nabla_{\hat T}z_i,\nabla_{\hat T}\hat T\rangle\\
&=&-\alpha^{-2}\hat  T(\alpha)z_i(\alpha)+\alpha^{-1}\hat T(z_i(\alpha))+\\
& &+\langle [z_i,\hat T],\nabla_{\hat T}\hat T\rangle-\langle \nabla_{z_i}\hat T,\nabla_{\hat T}\hat T\rangle\\
\langle \hat T,\nabla_{z_i}\nabla_{\hat T}\hat T\rangle&=&-\langle \nabla_{z_i}\hat T,\nabla_{\hat T}\hat T\rangle\\
\langle \hat T,\nabla_{\hat T}\nabla_{z_i}\hat T\rangle&=&-\langle \nabla_{\hat T}\hat T,\nabla_{z_i}\hat T\rangle\\
\langle \hat T,\nabla_{z_i}\nabla_{z_j}\hat T\rangle&=&-\langle \nabla_{z_i}\hat T,\nabla_{z_j}\hat T\rangle\\
\langle z_i,\nabla_{z_j}\nabla_{\hat T}\hat T\rangle&=& 
z_j \langle z_i,\nabla_{\hat T}\hat T\rangle 
- \langle \nabla_{z_j} z_i , \nabla_{\hat T}\hat T\rangle \\
\langle z_i,\nabla_{\hat T}\nabla_{z_j}\hat T\rangle&=&\nabla_{\hat T}\langle z_i,\nabla_{z_j}\hat T\rangle-\langle \nabla_{\hat T}z_i,\nabla_{z_j}\hat T\rangle\\
&=&\nabla_{\hat T}\langle z_i,\nabla_{z_j}\hat T\rangle+\langle [z_i,\hat T],\nabla_{z_j}\hat T\rangle-\langle \nabla_{z_i}\hat T,\nabla_{z_j}\hat T\rangle\\
\langle z_i,\nabla_{z_k}\nabla_{z_j}\hat T\rangle&=&\nabla_{z_k}\langle z_i,\nabla_{z_j}\hat T\rangle-\langle \nabla_{z_k}z_i,\nabla_{z_j}\hat T\rangle\\
&=&\nabla_{z_k}\langle z_i,\nabla_{z_j}\hat T\rangle-\langle D_{z_k}z_i,\nabla_{z_j}\hat T\rangle.
\end{eqnarray*}
By Lemma \ref{lem:fol1b}, Lemma \ref{lem:comut-dk} and the inequality \eqref{eq:nablaTest} 
and because the function $\epsilon_1$  which appears in the induced metric has 
bounded second derivative, we can estimate all but  the  last three components. 
They can be estimated if we show that 
$\nabla_{\hat T}\langle z_i,\nabla_{z_j}\hat T\rangle$ and  
$\nabla_{z_k}\langle z_i,\nabla_{z_j}\hat T\rangle$ are bounded.

Using the estimates for $\hat T$ in the proof of Lemma \ref{lem:fol1b}, 
the estimates for $k_{ij}$ in Lemma \ref{lem:kt} and the estimates for the
Christoffel symbols in \eqref{eq:delga1} and \eqref{eq:delgaA} one finds after a
lenghty calculation that  
$$
| \hat T( k(z_i,z_j) ) | \leq C 
$$
which is gives the required estimes for 
$\nabla_{\hat T}\langle z_i,\nabla_{z_j}\hat T\rangle$.

To estimate $\nabla_{z_k}\langle z_i,\nabla_{z_j}\hat T\rangle$, we
calculate 
the tangential  derivative of $k_{ij}$ from the tangential derivative of the unphysical $\tilde k_{ij}$ as follows.
\begin{eqnarray*}
\nabla_{v_l} k_{ij}&=& \nabla_{v_l}(\Omega^{-1}\tilde k_{ij}-\lambda\gamma_{ij})\\
&=&\Omega^{-2}v_l(\Omega)\tilde k_{ij}-\Omega^{-1}\nabla_{v_l}\tilde k_{ij}-v_l(\lambda)\gamma_{ij}-\lambda\nabla_{v_l}\gamma_{ij}\\
&=&\Omega^{-2}v_l(\Omega)\tilde k_{ij}-\Omega^{-1}\tilde \nabla_{v_l}\tilde k_{ij}-\Omega^{-2}C^l_{ij}\tilde k_{ij}-v_l(\lambda)\gamma_{ij}-\lambda k_{il}\hat T_j-\lambda k_{il}\hat T_j.
\end{eqnarray*}

Once again because the second derivatives of the functions $\epsilon_1$ and $\epsilon_2$ which appear in the induced metric and the unphysical second fundamental form are bounded we can estimate $\nabla_{v_l}\tilde k_{ij}$ and $v_l(\lambda) $. From this we obtain the desired  estimate of $\nabla_{z_k}\langle z_i,\nabla_{z_j}\hat T\rangle$.
\end{pf}

\begin{remark} \label{rem:VE-remark}
By Lemma \ref{lem:Hprop}, we have for $\epsilon_-<\tau <\tau_0^-$ 
$$
\partial_x H =\frac{1}{2}\tau^2(1+{\cal O}(x)),
$$
which implies that  in $V_E^-$ we can prove 
results similar to  Lemma \ref{lem:fol1b}, Lemma \ref{lem:kt} and Lemma
\ref{lem:T2} under much weaker conditions in the decay of the metric than those of the Definition \ref{def:asymp-schwarz}.
\end{remark}

The following Lemma gives a lower bound for the height function of a
spacelike hypersurface in $V_E^+$, 
assuming a lower bound at $x = x_0$. The proof uses an estimate for the
backward lightcones.
\begin{lemma}
\label{lem:lctau}
Let $(V,g)$ be asymptotically Schwarzschild. Then $\tau_0^+$ can be chosen
sufficiently large so that the following holds. 
Let $S = S_{\tau_0^+}$ and 
suppose that $M^+ $ is a compact hypersurface with height function 
$w = \tau \big{|}_{M} - \tau_0^+$  over $V_E^+\cap S$ satisfying 
$w(p)\geq \tau_+ - \tau_0^+$ when $x(p)=x_0$. Then if $\tau_+$
is sufficiently large, the lower bound 
\begin{equation}\label{eq:w-lower}
w(c(x))\geq \tau_+\sqrt{1-\frac{2}{x\tau_+}}-\tau_0^+
\end{equation}
holds for $x_0\geq x(p)\geq x_1$, where $c(x)$ is an integral curve 
\mnote{rewrite here, check this!}
\mnote{ the estimate is OK but I dont understand how  to rewrite this}
of $\hat v_1=v_1/|v_1|$ and $x_1$ is the root of 
the equation $\tau_+\sqrt{1-\frac{2}{x\tau_+}}-\tau_0^+ =0$. 
\end{lemma}
\begin{pf}
For motivation, consider the foliation in the Minkowski spacetime given by
$$
\tau^2=t^2-r^2 ,\quad \tau > 0.
$$
The height function of the backward light cone at $\tau = \tau_+$ with
respect to the slice $\tau = \tau_0^+$ satisfies
$$
v(x)\geq \tau_+\sqrt{1-\frac{2}{x\tau_+}}-\tau_0^+\geq 0
$$
when $2/\tau_+\approx x_1\leq x\leq x_0$. 

Now we consider the general case. Let $w$ be the height function of a spacelike hypersurface. Then it
satisfies
\begin{equation}
\label{eq:olikhet}
(\alpha-|\beta|)|Dw| \leq  1.
\end{equation}

Let $c$ be a curve with $\dot c = v_1/|V_1|$. Then 
\begin{eqnarray*}
\frac{d w(c)}{dt}&=& \langle Dw,\dot{c}\rangle\\
&=&\frac{w_x}{\sqrt{\langle v_1,v_1\rangle}}\\
&\leq&|Dw|,
\end{eqnarray*}
and inserting this in \eqref{eq:olikhet}, we obtain
\begin{equation}
f(w(x), x, y^A)w_x\leq 1.
\label{eq:dfg}
\end{equation}
where
\begin{equation}
f(w(x), x, y^A)=(\alpha-|\beta|)\frac{1}{\sqrt{\langle v_1, v_1\rangle}}.
\label{eq:fg}
\end{equation}

By Lemma \ref{lem:fol1b},  spacetimes which are asymptotically Schwarzschild 
satisfy
\begin{eqnarray*}
\alpha&=&\frac{1}{\sqrt{1+\epsilon_2}}+{\cal O}(x^2)\\
\beta^1&=&\frac{x}{\tau\sqrt{(1+\epsilon_1)}}+{\cal O}(x^2)\\
\sqrt{\gamma^{11}}&=&\frac{x}{\tau}
\sqrt {\frac{(1+\tau^2x^2)}{(1+\epsilon_1)}}+{\cal O}(x^2).
\end{eqnarray*}

Using this in Equation \eqref{eq:fg}, we obtain
$$
f(\tau,x,y^A)= \frac{x}{\tau\sqrt{(1+\epsilon_1)(1+\epsilon_2)}}\big ( \sqrt{1+\tau^2x^2}- \sqrt{1+\epsilon_2}+{\cal O}(x)\big ).
$$
Hence the differential inequality \eqref{eq:dfg} becomes
$$
 ( f_0(w(x),x,y^A)+f_1(w(x),x,y^A) )xw_x\leq w(x)+\tau_0^+
$$
where $f_0$ is given by 
$$
f_0(w(x),x,y^A)=\sqrt {1+(w(x)+\tau_0^+)^2x^2}-1.
$$
If $\tau_0^+$ is chosen sufficiently large, then $|f_1|<\epsilon\ll 1$. 
The differential equation
$$
\big(\sqrt {1+(w(x)+\tau_0^+)^2x^2}-1+\epsilon \big )xw'(x)= w(x)+\tau_0^+,
$$
has the solution
$$
\bigg ( \frac{y}{\big(\sqrt{y^2+1}+1\big)
\big(\sqrt{y^2+1}+\epsilon\big)}\bigg )^{\epsilon}
\big(\sqrt{y^2+1}+1\big) = Ax^{1+\epsilon},
$$
with $y=(w(x)+\tau_0^+)x$ and initial condition $x=x_0,\; y=\tau_+x_0$. 
Using this we get the estimate
$$
\sqrt{y^2+1}+1\geq \tau_+x.
$$
Hence a  lower bound of the height function $w$ along $x$ is given by
\mnote{why smash[b]??}
$$
w(c(x)) \geq \tau_+\sqrt{1-\frac{2}{\smash[b]{x\tau_+}}}- \tau_0^+
$$
for $x_0\geq x(p)\geq x_1\approx2/\tau_+$ and $\tau_+$ sufficiently large. 
\end{pf}
We are now ready to find the setup where the existence proof in Section
\ref{sec:existence} can be carried out. 
\begin{theorem} 
\label{thm:setup}
Let $(V,g)$ be an asymptotically Schwarzschild spacetime. Then there exist 
$\tau_0^-, \tau_0^+$, $\epsilon_-\leq \tau_0^- <  \tau_0^+ < \infty$ and $x_0 > 0$ so
that with $V_E^-$ and $V_E^+$ defined as in \eqref{eq:ve} with the given $x_0$
used in defining $V_E$ in Definition \ref{def:asymp-schwarz} so that the
following holds.
\begin{enumerate}
\item \label{point:kV+} 
In $V_E^+$, $S_{\tau}$ 
has mean curvature  $k<3$, the foliation $\{S_\tau\}$ satisfies \eqref{eq:max} and \eqref{eq:Ricci}, and the lower bound on the height function $w$ given by 
\eqref{eq:w-lower} holds. 
\mnote{\lars added ref to \eqref{w-lower}, slight rewrite}
\item \label{point:kV-} In $V_E^-$ , $S_{\tau}$ 
has mean curvature  $k>3$ and the foliation $\{S_\tau\}$ satisfies 
\eqref{eq:max} and  \eqref{eq:Ricci}.
\end{enumerate}
\end{theorem}
\begin{pf}
Point \ref{point:kV+} is a direct
consequence of Lemmata \ref{lem:fol1b}, \ref{lem:kt}, \ref{lem:T2} and 
\ref{lem:lctau}.
\mnote{\lars added ref to Lemma \ref{lem:lctau}}
By Lemma 
\ref{lem:Hprop}, we can write for $\epsilon_-\tau <\tau_0^-$ 
$$
\partial_x H =\frac{1}{2}\tau^2(1+{\cal O}(x)),
$$
which implies that  in $V_E^-$ we can prove 
results similar to  Lemmata \ref{lem:fol1b}, \ref{lem:kt} and 
\ref{lem:T2} under much weaker conditions in the decay of the metric than 
those of the Definition \ref{def:asymp-schwarz}.
Point \ref{point:kV-} follows. 
\end{pf}

This shows that in $V_E^+$ and $V_E^-$,  $\{S_{\tau}\}$ can be used as local 
upper respectively lower barriers for
CMC hypersurfaces with mean curvature $k=3$ and that the gradient estimates 
\cite[Theorem 3.1]{Bartnik} can be applied.

\section{Proof of existence of CMC hypersurfaces}
\label{sec:existence}
The main difficulty in proving the existence of CMC surfaces is in obtaining 
an apriori height bound. This can be done for spacetimes which are 
asymptotically Schwarzschild  and satisfy the following interior condition on 
the causal structure. 
\begin{definition}
\label{def:hypprop}
Let $(V,g)$ be an asymptotically Schwarzschild spacetime and let 
$0 < \tau_0^- < \tau_0^+ < \infty$ be given.
We say that $(V,g)$ satisfies the future interior condition if 
\begin{enumerate}
\item There exists a 
family of complete spacelike hypersurfaces 
$\{S_{\tau} \}_{\tau \in (0,\infty)}$, 
such that 
\begin{enumerate}
\item In the exterior region $V_E$, $S_{\tau}$ is as in Lemma \ref{lem:fol1b}.
\item $\{S_{\tau} \}_{\tau \in[\tau_0^+,\infty)}$ is a foliation of 
$D^+(S_{\tau_0^+})$.
\end{enumerate}
\item Let $V_I^+=D^+(S_{\tau^+_0}) \cap V_I$
denote the future interior region. 
Then there is a constant $C$ such that for all 
$q \in V$ with $x(q) =x_0$,
\begin{equation}
 {\sup}_{ p\in  V^+_I \setminus I^+ (q)}   (  \tau(p) -  \tau(q)  ) \leq  C_I,
\label{eq:intcond}
\end{equation}
where $I^+(q)$ denotes the chronological future of $q$.
\item \label{point:D(K)}
$D(K)$ is compact for every compact $K \subset V$.
\end{enumerate}
\end{definition}

\begin{remark}
\begin{enumerate}
\item 
The  future interior condition is satisfied in Minkowski space 
with $\{ S_{\tau} \}$ given by the  hyperboloids
$$
t^2-r^2 = \tau^2, \quad \tau\in (\tau_0,\infty), t > 0,
$$
for some $\tau_0 > 0$.
\item

The compactness of the domain of dependence of compact sets is 
fundamental in order to be able to apply the existence and regularity results 
for the Dirichlet problem for CMC hypersurfaces. 
\end{enumerate}
\end{remark}

In the rest of this section $(V,g)$ we will be assumed to be an asymptotically 
Schwarzschild spacetime, cf. Definition \ref{def:asymp-schwarz}, which 
satisfies the future interior condition introduced in Definition 
\ref{def:hypprop} and $\tau_0^-, \tau_0^+, x_0$ and  $V_E^+$ will be as in
Theorem \ref{thm:setup}. In the following also let $S = S_{\tau_0^+}$. 

The construction of the global upper barrier is based on a 
height estimate.  This is obtained using again the foliation given by 
Lemma \ref{lem:fol1b} and then applying a test function argument to the mean 
curvature formula \eqref{eq:mcf}.
\begin {theorem}
\label{thm:fund-est}
Suppose that $ M$ is a compact hypersurface with mean curvature $k=3$, such
that $\partial M\subset V_E\cap  S$ and let $w$ be the height function of 
$M^+ = M \cap D^+(S)$ over
$S$, defined by $w = \tau \big{|}_{M^+} - \tau_0^+$.
Then there is a  constant $C$ such that 
$\sup w\leq C$.
\end{theorem}
\begin{pf}
The outline of the proof is similar to that of \cite[Theorem 5.3]{Bartnik}. 
By Theorem \ref{thm:setup} the mean curvature $k$ of $S_{\tau}$ satisfies
$k<3$ in $V_E^+$ for $\tau > \tau_0^+$.
Therefore by the maximum principle (cf. eg. \cite[Chapter II]{protter}),
the height function $w$ has no local maximum in $\text{int} (V_E^+)$.
Thus we may assume that $w$ takes its supremum in $M^+ \cap V_I^+$.

Let  $M^+_E=  M \cap  V^+_E$. 
The boundary of $M^+_E$ is the union of components where $w=0$ and a set 
$B^+= M^+\cap B$ where $B = \{ p \in V_E : x(p) = x_0 \}$.

Let $\phi$ be a continuously  differentiable function on $\Re^+$, 
such that $\phi'(s) > 0, \phi(s)>0 $ for $s>0$. Using Equation \eqref{eq:mcf}, 
we have on $M^+$,
\begin{eqnarray}
\text{div}_{M}(\alpha^2\phi( w)\nabla^{M} w)&=&
\alpha\phi( w) (k\nu-\text{div}_M\hat T+\langle \nabla^{M} w,\nabla^{M} 
\alpha\rangle )+\nonumber\\
& &+(\nu^2-1)\phi'( w)\nonumber\\
&=&\phi( w) \bigg{(}
(\nu^2-1)\left ( \frac{\alpha k}{\nu+1}-\alpha k(\hat W, \hat W)
+\hat T(\alpha)+\frac{\phi '( w)}{\phi ( w)}\right )+\nonumber
 \\
& &  +\alpha \nu (k_M -k) \bigg{)} .
\label{eq:mcfphi}
\end{eqnarray}
Integrating \eqref{eq:mcfphi} by parts over $M^+_E$ gives
\begin {eqnarray}
\int_ {B^+}{\alpha}^2 \phi (w) \langle \nabla^{M} w, \sigma\rangle  d{\cal B^+}
&=& \int _{M^+_E}  \phi(w)( \nu ^2 - 1) \left (\frac {\phi '}{\phi} (w) +  {\hat T} 
( {\alpha})-  {\alpha}  k(\hat W, \hat W) +\nonumber \right. \\
& &\left. +\frac  {\alpha} { \nu + 1 }  {k} \right ) dv_{M^+_E}
+\int _{  {M^+_E}}  {\alpha} \phi (w) ( k_M -  k)  \nu  dv_{ M^+_E},
\label{eq:integral}
\end{eqnarray}
where  $\sigma$ is the outer normal of $B^+,\;dv_{ M^+_E} $ is the volume form on 
$ M^+_E$ and ${\cal B^+}$ is the induced area  on $B^+$.

Using Lemmata  \ref{lem:fol1b} and \ref{lem:kt} we estimate
\begin{equation}
\left | \hat T ( \alpha)- \alpha  k(\hat W, \hat W) +
\frac{\alpha}{ \nu + 1 }  k \right |
\leq \frac{1+\epsilon}{\tau}+(1-2\epsilon)x ,\quad  \text {in } M^+_E.
\label{eq:Ahatest}
\end{equation}

Let $\tau_+ -\tau_0^+ =\sup_{M^+}  w - C_I $. Using the interior 
condition  \eqref{eq:intcond}, we have that
$$
\inf_{p\in B^+}  w(p)\geq \tau_+-\tau_0^+ .
$$
Thus for all  $p\in B^+$,
$$
\tau_+ - \tau_0^+ + C_I \geq w(p) \geq \tau_+-\tau_0^+.
$$
Since we want a height estimate of $w$, we may suppose that 
$\tau_+-\tau_0^+>\frac{1}{x_0}$. Then  $  w\geq 1$ for all $w(p)\in B^+$.
The the gradient estimate \cite[Theorem 3.1]{Bartnik} applies and 
gives $ \nu (p)\leq C$ for all $p\in   B^+$, where $C$ is a constant does not 
depend on $\tau_+$. 
On the other hand, by assumption $V$ is asymptotically Schwarzschild  
and using \eqref{eq:gam1AB} we have
$$
d{\cal B^+} =\sqrt{{\rm det}g_{AB}(\tau)}dy^2dy^3 
\leq C \sqrt{{\rm det}g_{AB}(\tau^+_0)}dy^2 dy^3,
$$
where $g_{AB}(\tau)$ is the induced metric on $B^+$ and
$$
\langle \nabla ^{ M}  w, \sigma\rangle 
=|\nabla ^{ M}  w|\langle e_1, \sigma\rangle 
=\alpha^{-1}\sqrt{\nu^2-1} \langle e_1, \sigma\rangle.
$$
The boundary term in \eqref{eq:integral} is now  estimated by
\begin{equation}
\biggl|\int_ {B^+} {\alpha}^2 \phi ( w) \langle \nabla ^{ M}  w, \sigma\rangle d{\cal B^+}  \biggr|
\leq C\;\text{$\sup$}_{ M} \phi ( w).
\label{eq:term1}
\end{equation}
The volume form on $ M$ can be written \cite{Bartnik}
$$
dv_{M^+_E} = 
|1+ \beta\cdot D w|\nu ^{-1}\sqrt{{\rm det}\gamma(\tau)}dx dy^2 dy^3,
$$
but by \eqref{eq:gam1} we estimate in $V_E$
$$
\sqrt{{\rm det}\gamma(\tau)}\geq \sqrt{{\rm det}\gamma(\tau_0^+)}
$$
and
$$
|1+ \beta\cdot D w|\geq\epsilon>0 .
$$
Therefore considering $M^+_E$ as a graph  over 
$S' \subset  S \cap  V^+_E$ and using Theorem \ref{thm:setup}, 
we can estimate
\begin{equation}
\int_{M^+_E} {\alpha} \phi (w) ( k_M -  k)  \nu  dv_{ M^+_E}
\geq C\int_{ S' } \phi ( w) d\mu ( \gamma) .
\label{eq:term2i}
\end{equation}
Inserting \eqref{eq:Ahatest}, \eqref{eq:term1}, \eqref{eq:term2i} into \eqref{eq:integral} gives
\begin{eqnarray}
& &\int _{M^+_E}  \phi (  w)(\nu ^2 - 1) \bigg (\frac {\phi '}{\phi} (  w) 
- \frac{1+\epsilon}{ w+\tau_0}-(1-2\epsilon)x\bigg ) dv_{ M^+}+\nonumber\\
& &\quad+\int_ {S'}\phi (  w) d\mu ( \gamma ) \;
\leq C {\sup}_{M^+} \phi(w).
\label{eq:inequal}
\end{eqnarray}
We will  choose $\phi$ such that the first term of \eqref{eq:inequal} becomes 
positive and in consequence may be discarded, obtaining in this way
$$
\int_ {S'}\phi (  w) d\mu ( \gamma ) \leq C\;\text{$\sup$}_{ M} \phi (  w).
$$
\begin{figure}
\centering
\setlength{\unitlength}{0.0060in}
\begin{picture}(700,525)(0,-10)
\path(0,485)(0,485)(0,485)
        (0,485)(0,485)
\dashline{4.000}(135,465)(145,405)(155,355)
        (160,320)(165,275)(165,225)
        (165,170)(155,75)(145,30)
\dashline{4.000}(170,295)(280,300)(330,310)
        (395,320)(435,335)(490,355)(570,400)
\path(145,35)(560,340)(565,345)
\path(700,0)(700,510)(20,510)
        (20,0)(700,0)
\dashline{4.000}(165,265)(210,265)(235,270)
        (285,280)(395,305)(485,340)(565,385)
\path(455,125)(400,80)
\path(404.925,86.614)(400.000,80.000)(407.458,83.518)
\dashline{4.000}(285,175)(285,135)(285,90)(275,35)
\drawline(180,40)(180,40)
\dashline{4.000}(165,250)(180,265)(185,265)
\dashline{4.000}(165,230)(200,265)
\dashline{4.000}(165,210)(225,270)
\dashline{4.000}(165,180)(250,270)
\drawline(165,165)(165,165)
\dashline{4.000}(165,160)(280,275)
\dashline{4.000}(170,135)(315,285)
\dashline{4.000}(210,150)(345,295)
\dashline{4.000}(245,165)(365,295)
\dashline{4.000}(275,175)(395,305)
\dashline{4.000}(315,190)(425,315)
\dashline{4.000}(370,225)(455,325)
\dashline{4.000}(415,250)(490,340)(490,345)
\dashline{4.000}(475,295)(530,365)
\dashline{4.000}(535,340)(570,385)
\path(165,295)(455,15)
\thicklines
\spline(160,320)
(210,285)(235,250)
        (260,300)(305,235)(335,240)(355,210)
\spline(395,240)
(415,330)(430,355)
        (470,340)(535,340)
\thinlines
\spline(545,260)
(500,245)(475,265)
\path(482.496,261.564)(475.000,265.000)(479.998,258.441)
\thicklines
\spline(160,130)
(300,180)(410,245)(565,360)
\thinlines
\spline(580,335)
(550,320)(530,330)
\path(538.050,328.211)(530.000,330.000)(536.261,324.633)
\put(580,385){\makebox(0,0)[lb]{\raisebox{0pt}[0pt][0pt]{\shortstack[l]{{\footnotesize $\tau=\tau_+-1$}}}}}
\put(305,255){\makebox(0,0)[lb]{\raisebox{0pt}[0pt][0pt]{\shortstack[l]{{\footnotesize $M^+$}}}}}
\put(430,135){\makebox(0,0)[lb]{\raisebox{0pt}[0pt][0pt]{\shortstack[l]{{\footnotesize $\text{\footnotesize backward  lightcone}$}}}}}
\put(510,275){\makebox(0,0)[lb]{\raisebox{0pt}[0pt][0pt]{\shortstack[l]{{\footnotesize $\text{\footnotesize forward  lightcone}$}}}}}
\put(570,415){\makebox(0,0)[lb]{\raisebox{0pt}[0pt][0pt]{\shortstack[l]{{\footnotesize $\tau=\tau_+$}}}}}
\put(265,20){\makebox(0,0)[lb]{\raisebox{0pt}[0pt][0pt]{\shortstack[l]{{\footnotesize $x\approx \frac{2}{\tau_+}$}}}}}
\put(95,15){\makebox(0,0)[lb]{\raisebox{0pt}[0pt][0pt]{\shortstack[l]{{\footnotesize $x=x_0$}}}}}
\put(595,335){\makebox(0,0)[lb]{\raisebox{0pt}[0pt][0pt]{\shortstack[l]{{\footnotesize $S'$}}}}}
\end{picture}
\caption[]{$M^+$ and the backward light cone}
\label{fig:back}
\end{figure}
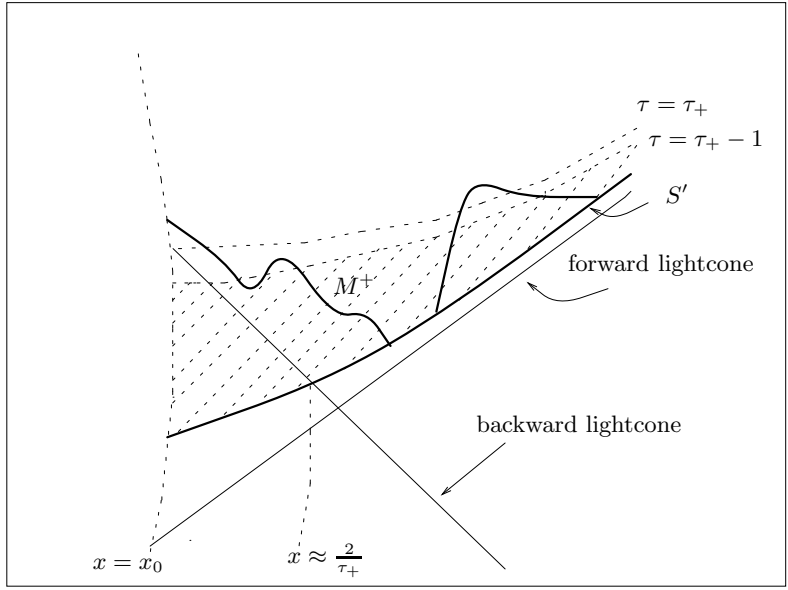
On the other hand we want to find $\phi$ such that $\sup_M w$ is 
bounded, in other words so that $\tau_+$ is bounded. 
In order to do this, observe that by Lemma \ref{lem:lctau}, 
(see Figure \ref{fig:back})
$$
w+\tau_0\geq\tau_+\sqrt{1-\frac{2}{\smash[b]{x(p)\tau_+}}}\; ,
$$
for $ 2/\tau_+\leq x(p)\leq x_0$, or
\begin{equation}
x\leq\frac{2}{\tau_+(1-((w+\tau_0)/\tau_+)^2) },\quad\text{for } \tau_0\leq
w+\tau_0\leq \tau_+-1 .
\label{eq:xup}
\end{equation}
To construct $\phi$ such that  the first integral 
in \eqref{eq:inequal} is positive, we solve the following differential equation
\begin{equation}
\frac{\phi '}{\phi}(s)= \frac{1+\epsilon}{ s+\tau_0}
+\frac{2(1-2\epsilon)}{\tau_+(1-((s+\tau_0)/\tau_+)^2) }.
\label{eq:testf}
\end{equation}
The solution is
$$
\phi (s)=\bigg(\frac{\tau_++\tau_0+s}{\tau_+-\tau_0-s}\bigg)^{1-2\epsilon}\; (s+\tau_0)^{1+\epsilon}.
$$
It is easily seen using \eqref{eq:xup} that this satisfies the required 
condition in the region $s \leq \tau_+-\tau_0-1$. In order to get a function 
which satisfies $\phi(0)=0$ and is defined for all $s\in \Re^+$, 
positive and continuously differentiable,  we construct $\phi$ as follows:
\[\phi (s) =  \left\{ \begin{array}{lllll}
                       D_1\; s^{D_2}      & \mbox{$0\leq s\leq 1$}\\
                                        &\\
                       \bigg(\dfrac{\tau_++\tau_0+s}{\tau_+-\tau_0-s}\bigg)^{1-2\epsilon}\; (s+\tau_0)^{1+\epsilon} & \mbox{$1\leq s \leq \tau_+-\tau_0-1$}\\
                                        &\\
                       (2\tau_+-1)^{1-2\epsilon}(\tau_+-1)^{1+\epsilon}\; \text{ e}^{D_3(s+\tau_0-\tau_++2)} & \mbox{$\tau_+-\tau_0-1\leq s,$}
                                                     \end{array}
               \right. \]
where
\begin{eqnarray*}
D_1&=& \bigg (\frac{\tau_++\tau_0+1}{\tau_+-\tau_0-1}\bigg )^{1-2\epsilon}\; 
(1+\tau_0)^{1+\epsilon}, \\
D_2&=&\frac{1+\epsilon}{1+\tau_0}+\frac{2(1-2\epsilon)}{\tau_+ (1-((1+\tau_0)/\tau_+)^2 )} , \\
D_3&=& \frac{1+\epsilon}{\tau_+-1}+\frac{1-2\epsilon}{1-1/2\tau_+}.
\end{eqnarray*}
With $\phi$ chosen as above, the first term in \eqref{eq:inequal} is 
positive (cf. \cite[Theorem 5.3]{iriondo} for details) and may be discarded, 
so \eqref{eq:inequal} becomes
\begin{equation}
\int_{ S'}\phi(w) d\mu ( \gamma) \leq C (\tau_+)^{2-\epsilon}.
\label{eq:uint}
\end{equation}
By construction $\phi(w)\geq 1$ when $w\geq 1$. By Lemma \ref{lem:lctau}, 
$w \geq 1$ when
$$
\frac{2(1+\delta )}{\tau_+}\leq x\leq x_0,
$$
where
$$
1+\delta =\frac{1}{1-((1+\tau_0)/\tau_+)^2}.
$$
Thus, from  the inequality \eqref{eq:uint} we get 
$$
\int_{2(1+\delta )/\tau_+}^{x_0} x^{-3}\int_{S^2}d\mu ( \tilde \gamma) \leq C (\tau_+)^{2-\epsilon}
$$
and integrating shows that $\tau_+\leq C$. 
\end{pf}

The main existence theorem follows from this estimate.
\begin{theorem}\label{thm:main}
There is a complete CMC hypersurface $ M$, $C^\infty$ 
on each compact set, with $k_M = 3$ and with the same cut at $\Scri ^+$ as
$S$. Furthermore there is a constant $C$ such that in $V_E$, 
$\nu=-\langle \hat T,N\rangle <C$, 
where $\hat T$and $ N$ are the normals of $S_{\tau}$ and $M$ respectively.
\end{theorem}
\begin{pf}
Consider the region of the spacetime given by
$$
V'= \{p \in V:p\in D^-(S_{\tau_{\max}}) \},
$$
where $\tau_{\max}$ is the height estimate given by Theorem
\ref{thm:fund-est}, see Figure \ref{fig:arzela}. 
\begin{figure}
\centering
\setlength{\unitlength}{0.0070in}
\begin{picture}(480,445)(0,-10)
\path(60,325)(240,145)
\path(240,145)(420,325)
\dashline{4.000}(120,405)(120,65)
\dashline{4.000}(360,405)(360,65)
\path(390,365)(380,335)
\path(380.632,343.222)(380.000,335.000)(384.427,341.957)
\path(480,0)(480,430)(0,430)
        (0,0)(480,0)
\path(45,295)(75,320)
\path(70.135,313.342)(75.000,320.000)(67.574,316.415)
\path(80,375)(95,350)
\path(89.169,355.831)(95.000,350.000)(92.599,357.889)
\path(90,250)(115,270)
\path(110.002,263.441)(115.000,270.000)(107.504,266.564)
\path(435,295)(410,325)
\path(416.658,320.135)(410.000,325.000)(413.585,317.574)
\dashline{4.000}(70,350)(60,340)
\dashline{4.000}(85,350)(60,325)
\dashline{4.000}(95,345)(70,320)
\dashline{4.000}(110,345)(75,310)
\dashline{4.000}(125,340)(85,300)
\dashline{4.000}(140,335)(95,290)
\dashline{4.000}(160,330)(105,275)
\dashline{4.000}(175,325)(115,265)
\dashline{4.000}(195,325)(130,260)
\dashline{4.000}(215,320)(140,245)
\dashline{4.000}(240,320)(155,235)
\dashline{4.000}(265,320)(170,220)
\dashline{4.000}(290,325)(180,205)
\dashline{4.000}(310,325)(195,195)
\dashline{4.000}(330,330)(205,180)
\dashline{4.000}(355,335)(220,170)
\dashline{4.000}(375,340)(230,160)
\dashline{4.000}(400,350)(240,145)
\dashline{4.000}(415,355)(340,245)
\dashline{4.000}(420,340)(395,300)
\spline(60,355)
(205,320)(305,320)(420,355)
\spline(100,300)
(145,290)(145,260)
        (185,210)(210,280)(240,235)
        (275,205)(295,210)(315,225)
        (330,250)(340,285)(370,315)(390,310)
\spline(60,335)
(165,245)(230,215)
        (260,215)(305,235)(375,295)(420,335)
\spline(85,315)
(110,310)(160,320)
        (180,245)(215,320)(245,270)
        (275,235)(300,240)(315,285)
        (340,315)(375,340)(410,325)
\spline(160,145)
(175,185)(225,195)
\path(217.548,191.470)(225.000,195.000)(216.763,195.392)
\put(375,370){\makebox(0,0)[lb]{\raisebox{0pt}[0pt][0pt]{\shortstack[l]{{\footnotesize $M_n$}}}}}
\put(10,280){\makebox(0,0)[lb]{\raisebox{0pt}[0pt][0pt]{\shortstack[l]{{\footnotesize $S_{\tau^+_0}$}}}}}
\put(420,280){\makebox(0,0)[lb]{\raisebox{0pt}[0pt][0pt]{\shortstack[l]{{\footnotesize $x=x_n$}}}}}
\put(60,385){\makebox(0,0)[lb]{\raisebox{0pt}[0pt][0pt]{\shortstack[l]{{\footnotesize $S_{\tau_{\max}}$}}}}}
\put(150,125){\makebox(0,0)[lb]{\raisebox{0pt}[0pt][0pt]{\shortstack[l]{{\footnotesize $V'$}}}}}
\put(35,235){\makebox(0,0)[lb]{\raisebox{0pt}[0pt][0pt]{\shortstack[l]{{\footnotesize $D^-(S_{\tau_{\max}})$}}}}}
\put(330,45){\makebox(0,0)[lb]{\raisebox{0pt}[0pt][0pt]{\shortstack[l]{{\footnotesize $x=x_K$}}}}}
\end{picture}
\caption{$M_n$ and $V'$}
\label{fig:arzela}
\end{figure}
Let $\{x_n\}_{n=1}^{\infty}$ be a decreasing sequence in $(0,x_0)$ such that 
$x_n \to 0$ as $n \to \infty$.
By Point \ref{point:D(K)} of Definition \ref{def:hypprop} and the fact
that the mean curvature equation $k(w)=3$ satisfies the mean curvature
structure condition with structure function $F \in C^{\infty}$, 
\cite[Theorem 4.1]{Bartnik2}
ensures that the Dirichlet problem
\begin{eqnarray*}
k_{M_n} &=& 3,\quad \text{in }  S_n = S\setminus 
S\cap\lbrace  p\in V : x(p) < x_n\rbrace \nonumber \\
\partial M_n &=& \partial S_n
\end{eqnarray*}
has solution $M_n$ which is a $C^{\infty}$ regular spacelike hypersurface
satisfying
$$
M_n \subset  V'.
$$
Note that $\Sigma(S_n) = \emptyset$ (cf. \cite[(3.13)]{Bartnik2}) since 
$S$ is spacelike.

We will now prove that the sequence $\{M_n\}$ has a convergent subsequence.
\mnote{ This is the new proof}
Construct a compact set $K$ as follows. Let $x_K \leq x_0$ be given and
define compact sets $S_K$, $L_K$ by 
\begin{eqnarray*}
S_K &=&S_{\tau_{\max}}\cap \{p\in V': x(p) \geq x_K\}, \\
L_K &=& \{p\in V : x(p)=x_K \mbox{ and } 0\leq u\leq
H(x_K,\tau_{\max})\},
\end{eqnarray*}
where $H$ is the height function used in the definition of the foliation 
$\{S_{\tau} \}$, cf. Section \ref{sec:barriers}. Note that the ``past
\mnote{\lars added some explanation here}
boundary'' of $L_K$ is the intersection with $\{x = x_K\}$ of the backward 
light cone of the cut of $S$ with $\Scri^{+}$, and the ``future boundary''
of $L_K$ is the intersection of the surface $S_{\tau_{\max}}$ 
with $\{x = x_K\}$.

Defining ${\cal S}_K =L_K\cup S_K$,
\mnote{\lars some rewriting here}
we have ${\cal S}_K$ compact. Hence in view of Point \ref{point:D(K)} of 
\mnote{added some here}
Definition \ref{def:hypprop}, the set 
$$
K=D^- ({\cal S}_K)
$$
is compact. 
Let now
$$
M^K_n =  M_n\cap K . 
$$
By construction of $M_n$, it follows using 
the height estimate from Theorem \ref{thm:fund-est} and causality, 
\mnote{\lars added a sentence here}
that $\partial M^K_n \subset L_K$. 
By the compactness of $K$ and using \cite[Theorem 3.8]{Bartnik2}, 
there is a subsequence $M^K_m$ and a weakly spacelike 
hypersurface  $M^K$ such that if $p$ is  an accumulation point  of $M^K_m$, 
$p\in M^K$ and $M^K-\Sigma(M^K) $ is a $C^\infty$ regular hypersurface with 
mean curvature $3$ where  $\Sigma(M^K)$ is the singular set consisting of a
union of null geodesics beginning and ending in $\partial M^K$. 

Because this holds for each $x_K \leq x_0$, by letting $x_K \to 0$,  we
find a hypersurface $M$ and a singular subset $\Sigma(M)$ such that 
$M - \Sigma(M)$ is a $C^{\infty}$ regular hypersurface with mean curvature
$3$ and $\Sigma(M)$ consists of a union of null geodesics beginning and
ending at $\Scri^+$. We claim that $\Sigma(M) = \emptyset$. 
To see this it is sufficient to recall that $\Scri^+$ is in the future of $V$
and noting that the existence of a null geodesic beginning and ending at
$\Scri^+$ would contradict this fact.

It follows that $M$ is $C^{\infty}$ regular with $k_M = 3$. 
By construction $M \subset D^-(S_{\tau_{\max}})$ and therefore has 
the same cut at $\Scri^+$ as $S_{\tau_{\max}}$ and hence the same cut as
$S$. This proves the existence of $M$.

\noindent{\bf Claim:} There is a constant $C$ such that in $V_E$,
\begin{equation}\label{eq:nuest}
\nu=-\langle \hat T,N\rangle <C .
\end{equation}

To prove the claim, we consider $M^-_n=M_n\cap V^-_E$ and 
$M^-=M\cap V^-_E$ as given 
by the height functions $w_n$ respectively $w$.

The $S_{\tau}$ for $\tau < \tau_0^-$ are lower barriers for CMC hypersurfaces
with $k=3$ and therefore the maximum principle argument applied as in  the
proof of Theorem \ref{thm:fund-est} implies that 
the infimum of  $w_n$  must be  attained  at a point $p$ 
satisfying $p \in M_n \cap B$ or $p\in \partial M_n$. 
We will use this to prove a lower bound for $w$.

By construction, if $p \in \partial M_n$, $w(p) = \tau_0^+$. 
It remains to consider the case when the infimum of $w_n$ is  
attained in $M_n \cap B$.
By the above, $w_n(p) \to w(p)$ and $M$ is strictly spacelike on compacts. 
As a consequence there exists $N_0 >0$ and $\tau_- \in (0, \tau_0^-)$ 
such that $w_n(p) \geq \tau_->0$, when 
$n\geq N_0$ and $p \in B$, which in turn implies that 
$ w_n(p) \geq \tau_-$, when $p\in V_E^-$. 
It follows that 
$$
w\geq \tau_-,\quad\text{on } V_E^-.
$$

Choose $\epsilon> 0$ in \eqref{eq:ve} such that $0<\epsilon\leq \tau_-$. By Point
\ref{point:kV-} of Theorem \ref{thm:setup}, 
the foliation $\{S_\tau\}$ satisfies \eqref{eq:max} and \eqref{eq:Ricci}
and therefore  the gradient estimate 
\cite[Corollary 3.4]{Bartnik} implies 
$$
\nu < C ,\quad\text{on } M 
$$
which completes the proof of the claim.

It remains to prove that $M$ is complete.
Let $M_E = M \cap V_E$ and let $p \in M_E$ be given.
Let $c: [0,\infty[ \, \to M_E $ be an arclength parametrized 
curve in $M_E$ with $c(0) =p$. 
To prove that $M$ is
complete, it is sufficient to prove that $x(c(t)) > 0$ for $t < \infty$,
since this implies that any geodesic in $M$ reaching infinity must have
infinite length.

Let $\sigma$ be the metric induced on $M$ from $g$.
Choose the adapted ON frame $\{e_i\}$ on $M$ as in Section
\ref{sec:prelim}. 
Let $\tilde e_i = \Omega^{-1} e_i$ be the corresponding ON frame 
w.r.t. the conformally related metric $\tilde \sigma = \Omega^2 \sigma$
and let $\tilde \nabla^M x = \tilde e_i(x) \tilde e_i$ be the gradient of $x$ 
in $M$ w.r.t. $\tilde \sigma$.
We compute
\begin{equation}
| \frac{d}{dt} x (c(t)) | \leq \Omega (c(t))  
\Vert \tilde \nabla^M x \Vert_{\tilde
\sigma} \Vert \dot c \Vert_{\sigma} 
= \Omega  (c(t)) \Vert \tilde \nabla^M x \Vert_{\tilde \sigma} ,
\label{eq:xestim}
\end{equation}
where we used the assumption that $c$ is arclength parametrized.
Using the $\nu$ estimate in 
\eqref{eq:nuest}, the definition of the adapted
frame and \eqref{eq:framerel} together with the fact that 
$x \in C^{\infty}(\tilde V)$, 
we find that $\Vert \tilde \nabla^M x \Vert_{\tilde \sigma}
\leq C$ in $V_E$ for some constant $C < \infty$. 
From \eqref{eq:xestim} and the assumptions on $\Omega$ we now get 
$$
| \frac{d}{dt} \ln( x(c(t))) | \leq C
$$
which after integrating gives 
$$
\frac{x(p)}{x(c(t))} \leq C' e^{Ct}
$$
which completes the proof.
\end{pf}
This result remains valid for hypersufaces with arbitrary constant mean curvature. We  have chosen $k=3$ for simplicity.

\appendix
\section{The estimate of $\delta \Gamma ^\mu_{\nu \lambda}$}
By definition the unphysical metric can be written as 
$$
(\tilde g_{\mu\nu})=\left(\begin{array}{cccc}
-x^2h(x)+{\cal O}(x^6)  \;& 1   & {\cal O}(x^3) &{\cal O}(x^3)\\
                           & 0          & 0         & 0\\
\\
&&\multicolumn{2}{c}{\smash{\raise6pt\hbox{$\tilde g_{AB}$}}}
         \end{array}     \right) 
$$
from which it follows that the inverse metric takes the form
$$
(\tilde g^{\mu\nu})=\left(\begin{array}{cccc}
    0 &  1     & 0         & 0\\
\; &x^2h(x)+{\cal O}(x^6)  \;   & {\cal O}(x^3) &{\cal O}(x^3)\\
\\
&&\multicolumn{2}{c}{\smash{\raise6pt\hbox{$\tilde g^{AB}$}}}
         \end{array}     \right). 
$$
>From this we can obtain the Christoffel symbols needes to 
compute $\tilde k_{11}$.

$\tilde \Gamma^\mu_{xx}$ are zero because $x$ is the affine parameter of 
null geodesics. The other components can easily be computed from the above
expression of the metric and are of the form
\begin{eqnarray}
\tilde \Gamma^u_{uu}&=&\tfrac{1}{2}\tilde g^{ux}(2\tilde g_{ux,u}-\tilde g_{uu,x})=\hat \Gamma^u_{uu}+{\cal O}(x^3), \nonumber\\
\tilde \Gamma^x_{ux}&=&\tfrac{1}{2}\tilde g^{ux}\tilde g_{uu,x}+\tfrac{1}{2}\tilde g^{xA}\tilde g_{Au,x}=\hat \Gamma^x_{xu}+{\cal O}(x^3),\quad \tilde \Gamma^u_{ux}=0, \nonumber\\
\tilde \Gamma^x_{uu}&=&\tfrac{1}{2}\tilde g^{ux}\tilde g_{uu,u}+\tfrac{1}{2}\tilde g^{xx}(2\tilde g_{ux,u}-\tilde g_{uu,x})+\tfrac{1}{2}\tilde g^{xA}(2\tilde g_{uA,u}-\tilde g_{uu,A}) \nonumber\\
&=&\hat \Gamma^x_{uu}+{\cal O}(x^5),
\label{eq:delga1}
\end{eqnarray}
where the 
$\hat \Gamma$ are the Christoffel symbols for the conformal Schwarzschild 
metric. 
The last equality requires that $\tilde g_{uu,u}={\cal O}(x^5)$.
In the other directions we need to  estimate
\begin{eqnarray}
\tilde \Gamma^u_{uA}&=&-\tfrac{1}{2}\tilde g^{xu}\tilde g_{uA,x}={\cal O}(x^2),
\nonumber\\
\tilde \Gamma^x_{xA}&=&\tfrac{1}{2}\tilde g^{ux}\tilde g_{uA,x}+\tfrac{1}{2}\tilde g^{xC}\tilde g_{AC,x}={\cal O}(x^2), 
\nonumber\\
\tilde\Gamma^x_{uA}&=&\tfrac{1}{2}\tilde g^{ux}\tilde g_{uu,A}-\tfrac{1}{2}\tilde g^{xx}\tilde g_{uA,x}+\tfrac{1}{2}\tilde g^{xC}(\tilde g_{uC,A}+\tilde g_{AC,u}-\tilde g_{uA,C})={\cal O}(x^4), \nonumber\\
\tilde \Gamma^u_{AB}&=&-\tfrac{1}{2}\tilde g^{ux}\tilde g_{AB,x}={\cal O}(1),\quad \tilde \Gamma^u_{xA}=0, \nonumber\\
\tilde \Gamma^x_{AB}&=&\tfrac{1}{2}\tilde g^{ux}(\tilde g_{uB,A}+\tilde g_{Au,B}-\tilde g_{AB,u})-\tfrac{1}{2}\tilde g^{xx}\tilde g_{AB,x}+\nonumber\\
& &+\tfrac{1}{2}\tilde g^{xC}(\tilde g_{BC,A}+\tilde g_{AC,B}-\tilde g_{AB,C})={\cal O}(x^2) .
\label{eq:delgaA}
\end{eqnarray}
Once again the last equality requires that $\tilde g_{AB,u}={\cal O}(x^2)$.


\end{document}